\documentclass[
 reprint,
amsmath,
amssymb,
aps,
pra,
]{revtex4-1}
\usepackage{comment}
\usepackage{graphicx}
\usepackage{dcolumn}
\usepackage{bm}
\usepackage{color}
\usepackage{multirow}
\usepackage{geometry}
\usepackage{multirow,array}

\newcolumntype{P}[1]{>{\centering\arraybackslash}p{#1}}
\begin{document}


\title{Critical analysis of the re-entrant localization transition in a one-dimensional dimerized quasiperiodic lattice}

\author{Shilpi Roy, Sourav Chattopadhyay, Tapan Mishra and  Saurabh Basu}
\affiliation{Department of Physics, Indian Institute of Technology Guwahati-Guwahati, 781039 Assam, India}

\date{\today}

\begin{abstract}
A re-entrant localization transition has been predicted recently in a one-dimensional quasiperiodic lattice with dimerized hopping between the nearest-neighbour sites (Phys. Rev. Lett. {\bf 126} 106803 (2021)) \cite{PhysRevLett.126.106803}. It has been shown that the interplay between the hopping dimerization and a staggered quasi-periodic disorder manifests two localization transitions through two intermediate phases  resulting in four critical points as a function of the quasiperiodic potential. In this paper, we study the phenomenon of the re-entrant localization transition by examining the spectral properties of the states. By performing a systematic finite-size scaling analysis for a fixed value of the hopping dimerization, we obtain accurate critical disorder strengths for different transitions and the associated critical exponents. 
Moreover, through a multifractal analysis, we study the critical nature of the states across the localization transitions by computing the mass exponents and the corresponding fractal dimensions of the states. 
Further, we complement the critical nature of the states by computing the Hausdorff dimensions. 
\end{abstract}

\maketitle


\section{\label{sec:level1} Introduction}
Anderson localization is a ubiquitous phenomenon in condensed matter that involves lattices with random on-site disorder~\cite{PhysRev.109.1492}. The phenomenon which is marked by the transition of all the extended/delocalized single particle states to localized states at a critical disorder strength is absent in one and two dimensions~\cite{PhysRevLett.42.673}. However, an intermediate between a periodic and a fully disordered systems namely the quasiperiodic lattices exhibit delocalization-localization (DL) transitions in low dimensions~\cite{SOKOLOFF1985189}. This remarkable property of the quasiperiodic lattices have encouraged the study of DL transitions in various different models~\cite{PhysRevLett.49.833, simon1985almost,PhysRevLett.51.1198,PhysRevLett.50.1873,aubry1980analyticity}. Due to the easier experimental access  over the random lattices, the quasiperiodic lattices have been created and studied in different experimental setups such as in optical lattices, photonic lattices, optical cavities, superconducting circuits etc.~\cite{PhysRevLett.103.013901, PhysRevLett.55.1768, PhysRevLett.112.146404,roushan2017spectroscopic}. These developments have facilitated the observations of interesting physical phenomena such as the Anderson localization~\cite{billy2008direct,aspect2009anderson,lahini2008anderson,PhysRevLett.120.160404},  Bose glasses~\cite{PhysRevLett.98.130404}, emergence of long-ranged periodic order~\cite{PhysRevA.72.053607,viebahn2019matter}, and many-body localization \cite{PhysRevB.87.134202,PhysRevLett.119.260401,PhysRevX.7.041047} etc.

Among the various quasiperiodic lattice models, the simplest, yet interesting model is the Aubry-Andr\'e model (AA) ~\cite{aubry1980analyticity} which exhibits a sharp DL transition. The transition occurs at an exact critical value of the quasiperiodic potential due to the self-dual nature of the AA model ~\cite{aubry1980analyticity,soukoulis1982localization,kohmoto1983metal}. The sharp DL transition dictates the absence of any energy-dependent mobility edge (ME) (the critical energy which separates the localized and delocalized states) at the transition. Hence, the system undergoes a transition from all states extended to all states localized through the critical point. However, the breaking of the self duality of the AA model or further generalizations of it have shown to exhibit the DL transition through intermediate/critical regions hosting the mobility edge e.g. zig-zag lattices \cite{an2018engineering}, flat-band lattices \cite{bodyfelt2014flatbands},  quasiperiodic mosaic lattices \cite{wang2020one}, shallow bichromatic potentials \cite{yao2019critical}, in presence of longer-range hopping~\cite{deng2019one}, and the generalized AA model \cite{PhysRevB.41.5544,PhysRevA.80.021603,PhysRevLett.104.070601,PhysRevB.83.075105,PhysRevLett.114.146601,PhysRevLett.126.040603,PhysRevB.103.184203} etc. Furthermore, quasiperiodic lattices exhibit a plethora of unique characteristics, including critical spectra, multifractal wavefunctions at and away from the critical points corresponding to the DL transition, and the existence of mixed phases hosting the likes of both the localized and the delocalized states which have been studied in great detail in various systems\cite{PhysRevLett.51.1198,Siebesma_1987,yao2019critical,deng2019one,PhysRevLett.50.1870,szabo2018non,PhysRevB.50.11365,PhysRevB.96.085119,10.21468/SciPostPhys.4.5.025,PhysRevB.34.2041,PhysRevB.35.1020}.

Recently, in the context of the quasiperiodic lattices, a re-entrant localization transition has been predicted by some of us in Ref.~\cite{PhysRevLett.126.106803}. It was shown that a one dimensional quasiperiodic model of AA type with dimerized hopping and staggered quasi-periodic disorder can undergo two localization transitions at the single particle level. In other words, for some specific dimerization strengths and as a function of disorder, the system first undergoes a localization transition where all the single particle states get localized. Further increase in the disorder strength turns some of the localized states extended, and eventually the system undergoes another localization transition at a larger disorder strength where all the single particle states get localized for the second time. Both the localization transitions are found to occur through two intermediate regions hosting the MEs resulting in four critical points as a function of the quasiperiodic potential strength. While the detailed phase diagram depicting the re-entrant localization transition associated to this model has been discussed in Ref.~\cite{PhysRevLett.126.106803}, a thorough understanding of the phase transitions can be unveiled via a quantitative analysis of the critical properties which is relevant and of topical interest.  

In this paper, we study the critical properties of the re-entrant localization transitions described above. By using appropriate scaling functions, we explore the critical points for different phase transitions. In our analysis, we are able to obtain the critical points, critical exponents and scaling behaviour associated to the first localization transition. However, near the second localization transition, the scaling behaviour is not well captured in our analysis. We further analyse the spectrum near the second localization transition and find the existence of multifractal states and identify the critical regimes.

The paper is organized as follows. First, we describe the model, approach and the associated phase diagram depicting the re-entrant localization transition briefly in Sec-II. In Sec-III we discuss the results in detail and finally in Sec-IV we provide a brief conclusion. 

\section{Model and approach}\label{model}
The Hamiltonian with hopping dimerization and staggered quasiperiodic disorder on a one-dimensional chain is written as ~\cite{PhysRevLett.126.106803}, 
\begin{align}
H=&-t_{1}\sum_{i=1}^{N}(c^{\dagger}_{i,B}c_{i, A}+{\rm {H.c.}})
-t_{2}\sum_{i=1}^{N-1}(c^{\dagger}_{i+1, A}c_{i, B}+ {\rm {H.c.}}) \nonumber\\
+&\sum_{i=1}^{N} \lambda_{A}n_{i,A}\cos[2\pi\beta(2i-1)+\phi] \nonumber \\
+&\sum_{i=1}^{N}\lambda_{B}n_{i,B}\cos[2\pi\beta(2i)+\phi]
\label{eqn:ham}
\end{align}
where $L=2N$ with $N$ being the number of unit cells that are denoted by the index $i$ and $L$ is the total system size. Here, a unit cell comprises of two sublattice sites, namely, $A$ and $B$ where the corresponding creation (annihilation) operators are denoted by  $c^{\dagger}_{i,A}~(c_{i, A})$ 
and $c^{\dagger}_{i, B}~(c_{i, B}$) respectively. The inter-cell hopping between the two sublattices is denoted by $t_{2}$, while $t_{1}$ refers to the intra-cell hopping. The hopping dimerization is introduced by defining $\delta=t_{2}/t_{1}$ and making $\delta\neq1$. We have taken $t_{1}$ as the unit of energy throughout the study.
The on-site quasiperiodic potential at the sublattice site $A~(B)$ is given by $\lambda_{A}$~($\lambda_B$). The quasiperiodicity is achieved by considering an irrational $\beta$. In particular, we take it as the inverse of the golden mean, namely $\beta=\frac{(\sqrt{5}-1)}{2}$ \cite{szabo2018non}. $\phi$ denotes the phase difference between the lattices that form the quasiperiodic lattice. In our studies, we consider very large system sizes $L$ up to a maximum of $35422$ sites for which $\phi$ can be set to zero without any loss of generality. 

\begin{figure}[htb!]
\centerline{\hfill
\includegraphics[width=0.5\textwidth]{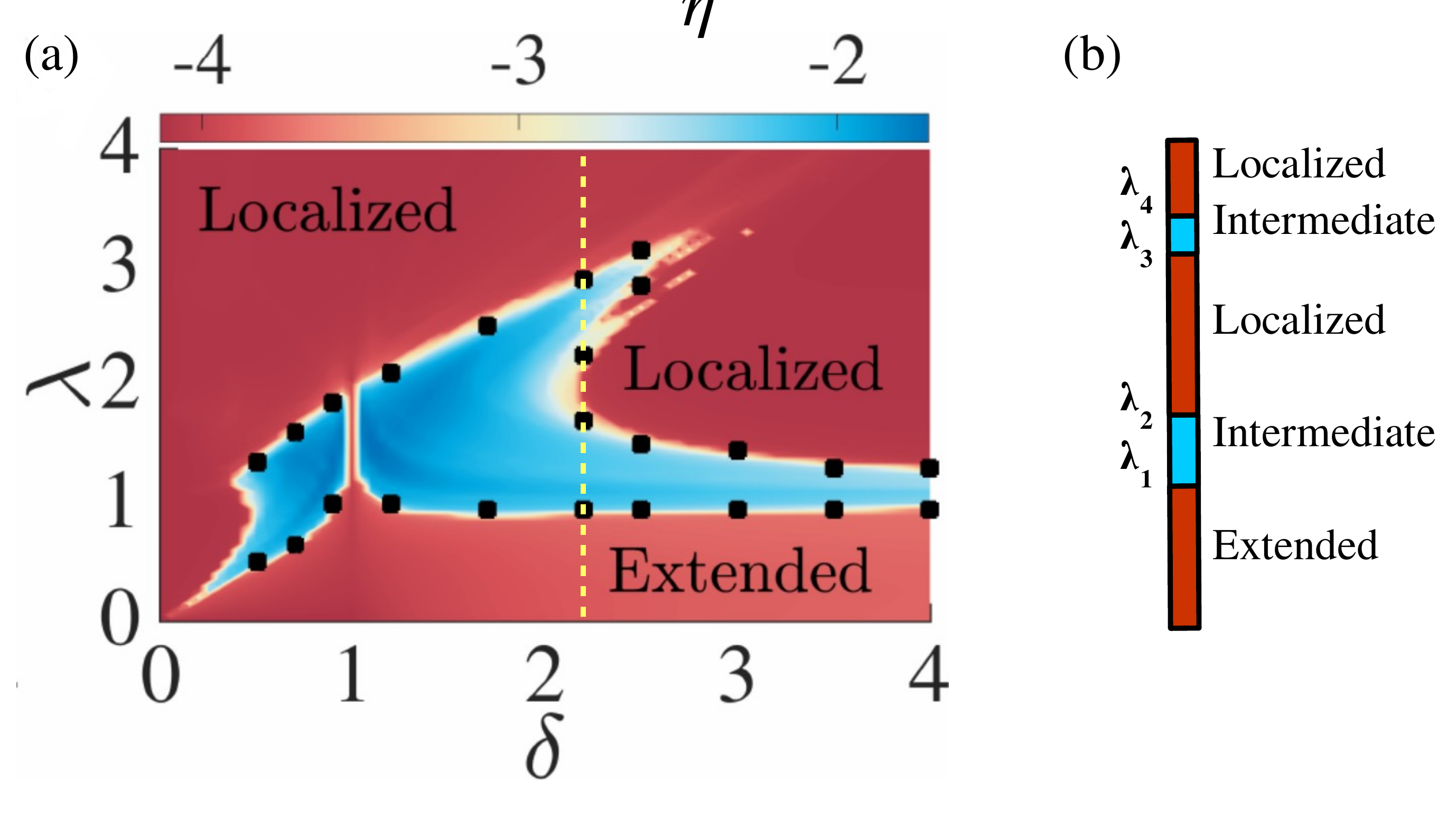}
\hfill}
\caption{Phase diagram is plotted as a function of hopping dimerization $\delta$ and disorder strength $\lambda$ in (a).  In (b), a schematic picture of the series of transitions is shown for $\delta=2.2$ (mark by the dashed line in (a)).\cite{PhysRevLett.126.106803}}
\label{fig:scheme1}
\end{figure}

The localization properties of the model shown in Eq.~\ref{eqn:ham} has been discussed in detail in Ref.~\cite{PhysRevLett.126.106803}. It has been shown that the system exhibits a re-entrant localization transition in the limit of staggered disorder i.e.  $\lambda_A=-\lambda_B=\lambda$, which has been depicted as a phase diagram in the $\delta$ - $\lambda$ plane in  Fig.~\ref{fig:scheme1} (see Ref.~\cite{PhysRevLett.126.106803} for details). It can be seen for Fig.~\ref{fig:scheme1}(a) that in certain values of $\delta$, the system undergoes two localization transitions through two intermediate phases as a function of $\lambda$ (e.g. the vertical dashed line in Fig.~\ref{fig:scheme1}(a)). This results in four critical points such as $\lambda_1$, $\lambda_2$, $\lambda_3$,  and $\lambda_4$ as schematically depicted in Fig.~\ref{fig:scheme1}(b). In this work, our focus is to explore the critical properties of such re-entrant phase transitions. 

We study the critical state behaviour through systematic finite-size scaling analysis following Ref.~\cite{Hashimoto_1992}. In general, localization in a disordered system can be characterized by using the normalized participation ratio, (NPR) which for the $m$-th eigenstate is defined as,
\begin{equation}
{\rm{NPR}}^{m}=\bigg[\it{L}~\sum_{i=1}^{\it{L}} {|\phi_{i}^{m}|^4} \bigg] ^{-1}=\frac{{\rm{PR}}^{m}}{\it{L}}
\end{equation} 
where $L$ is the system size and PR is the participation ratio.  We can identify the order parameter for the DL transition as,
\begin{equation}
\sigma^{m}=\sqrt{\frac{\rm{PR}^{m}}{L}}=\sqrt{\rm{NPR}^{m}}
\label{sigmaP}
\end{equation}
In the extended regime, $\rm{PR}$ grows linearly with system size $L$, while it vanishes in the localized regime in the thermodynamic limit. In the vicinity of the phase transition, the observables show power law behaviour with their critical exponents behaving as, \cite{Hashimoto_1992}
\begin{equation}\label{eq4}
\sigma \sim (-\varepsilon)^{\beta} \qquad 
{\rm{PR}} \sim \varepsilon^{-\gamma}  \qquad 
\xi\sim |\varepsilon|^{-\nu} .
\end{equation}
Here, $\varepsilon=(\lambda-\lambda_{c})/\lambda_{c}$ is the reduced disorder potential strength with $\lambda_{c}$ being the critical disorder strength for the DL transitions and $\xi$ is the correlation (or localization) length.
$\beta$, $\gamma$ and $\nu$ are the order parameter exponent, participation ratio exponent and the correlation length exponent respectively.

The critical point for the transitions $\lambda_{c}$ and the critical exponent ratio $\gamma/\nu$ are determined using the two system size-variable function $R~[L,L^{\prime}]$ given by, \cite{Hashimoto_1992}
\begin{equation}
R~[L,L^{\prime}]=\frac{log(\sigma^{2}_{L}/\sigma^{2}_{L^{\prime}})} {log(L/L^{\prime})}+1
\label{EqR}
\end{equation}
The variation of $R[L,L^{\prime}]$ with the strength of the potential in the vicinity of the critical point for several pairs of system of sizes $L$ and $L'$ intersect each other at a common fixed point. The critical potential strength $\lambda_{c}$ and the exponent ratio $\gamma/\nu$ are determined from the abscissa and the ordinate of the common crossing point respectively.

In the vicinity of the critical point, a finite-size scaling form of the order parameter $\sigma$ for finite system is defined by,
\begin{eqnarray}
\sigma = L^{-\beta/\nu}F(\varepsilon L^{1/\nu})
\label{beta}
\end{eqnarray}
where $F$ is a scaling function . Similarly, a finite-size scaling form of PR for a finite sized system is defined by,
\begin{eqnarray}
{\rm{PR}} = L^{\gamma/\nu}G(\varepsilon L^{1/\nu})
\label{gamma}
\end{eqnarray}
where $G$ is another scaling function. 
Using Eq.(\ref{sigmaP}), Eq.(\ref{gamma}) can be re-written as, 
\begin{equation}
 \sigma^2L = L^{\gamma/\nu}G(\varepsilon L^{1/\nu})\nonumber
\end{equation}
which can be further expressed as,
\begin{equation}
\sigma^2 = L^{\gamma/\nu-1}G(\varepsilon L^{1/\nu})
\label{gamma-1}
\end{equation}
Hence, a plot of $\sigma^2L^{1-\gamma/\nu}$ versus $\varepsilon L^{1/\nu}$ for different system sizes $L$ should fall onto a single curve denoted by  $G(\varepsilon L^{1/\nu})$ if the critical potential strength  $\lambda_{c}$ and the critical exponents are correctly determined. Note that $\sigma$ and PR here are the average values taken over the states considered. 

\section{Results}\label{res}

\subsection{Critical state analysis}\label{res1}

In this subsection we examine the re-entrant localization transition as depicted in the phase diagram of Fig.~\ref{fig:scheme1}(a). The red regions correspond to the extended or localized phases as denoted in the phase diagram and the central blue region bounded by the dark symbols is the intermediate phase. It can be seen from the phase diagram that for a range of $\delta$, the system undergoes two localization transitions as a function of $\lambda$ indicating the re-entrant localization transition.  Although, the re-entrant localization is feasible in both the regimes  of hopping dimerization corresponding to $\delta<1$ and $\delta >1$~\cite{PhysRevLett.126.106803}, for our discussion we restrict ourselves in the regime of $\delta>1$ for concreteness. For our analysis, we explore the critical properties for a cut through the phase diagram along the $y-$ axis at $\delta=2.2$ (dashed yellow line in Fig.~\ref{fig:scheme1}(a)). As $\lambda$ is increased, the system as a whole undergoes two localization transitions through two intermediate phases exhibiting a series of transitions from extended - intermediate - localized - intermediate - localized phases occurring at four critical points, $\lambda_1,~\lambda_2,~\lambda_3$ and $\lambda_4$ respectively. In the following our focus is to determine these critical points of transitions through finite-size scaling analysis. 
\begin{figure}[!t]
\centering
{\includegraphics[width=\linewidth]{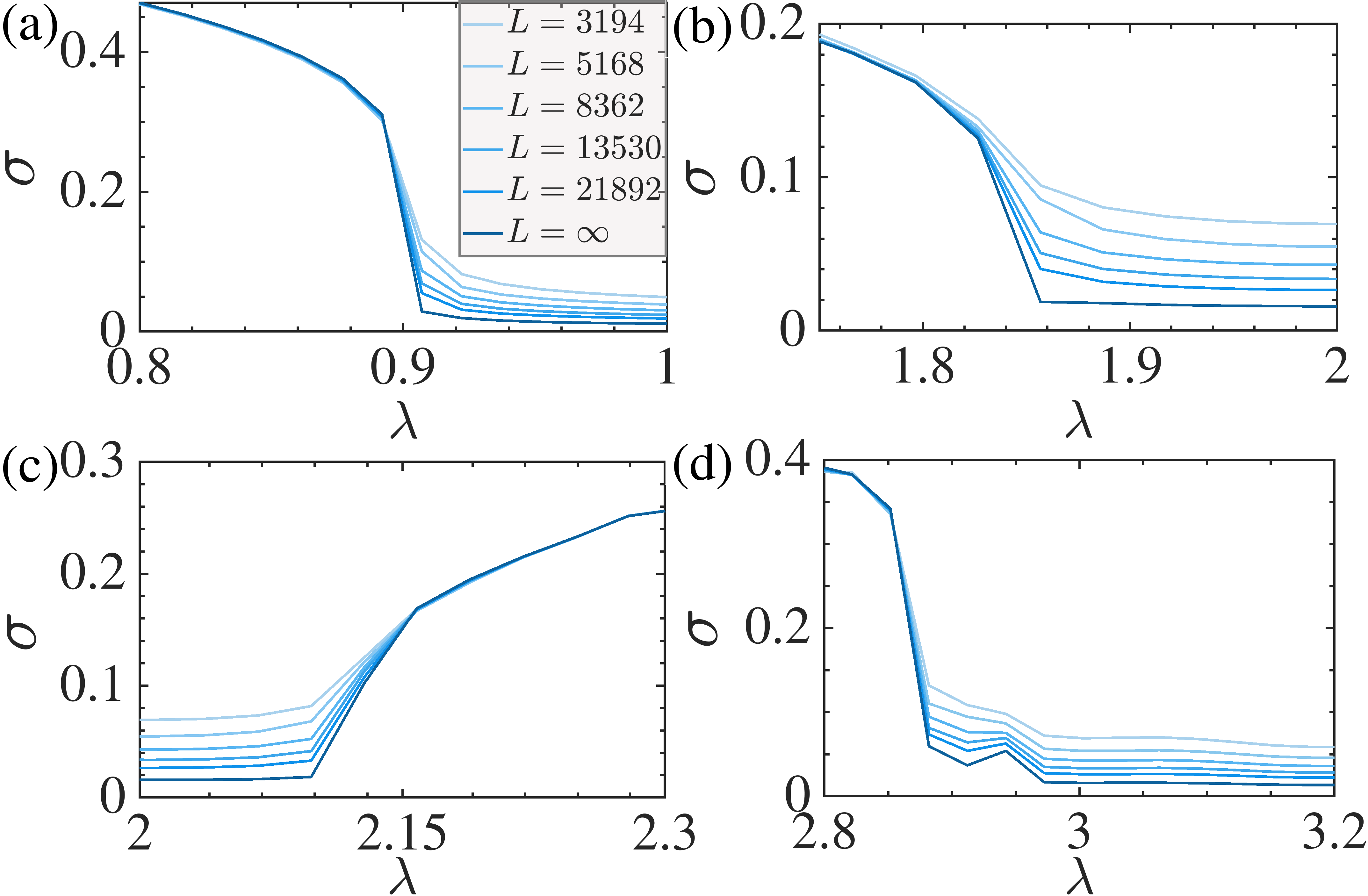}}
\caption{ The order parameter $\sigma$ is plotted as a function of $\lambda$ corresponding to four different critical transition points $\lambda_{1}$, $\lambda_{2}$, $\lambda_{3}$, and $\lambda_{4}$. We consider the states within a narrow band with indices (0-0.05), (0.45-0.5), (0.45-0.5) and (0.45-0.5) for the calculation of $\sigma$ in (a), (b), (c), and (d) respectively.  The color gradient in increasing order indicate different system sizes from small to large. The curve with deep blue color is obtained by using finite-size extrapolation. }
\label{fig:sigma_lambda}
\end{figure}

\begin{figure}[t]
\centering
{\includegraphics[width=\linewidth]{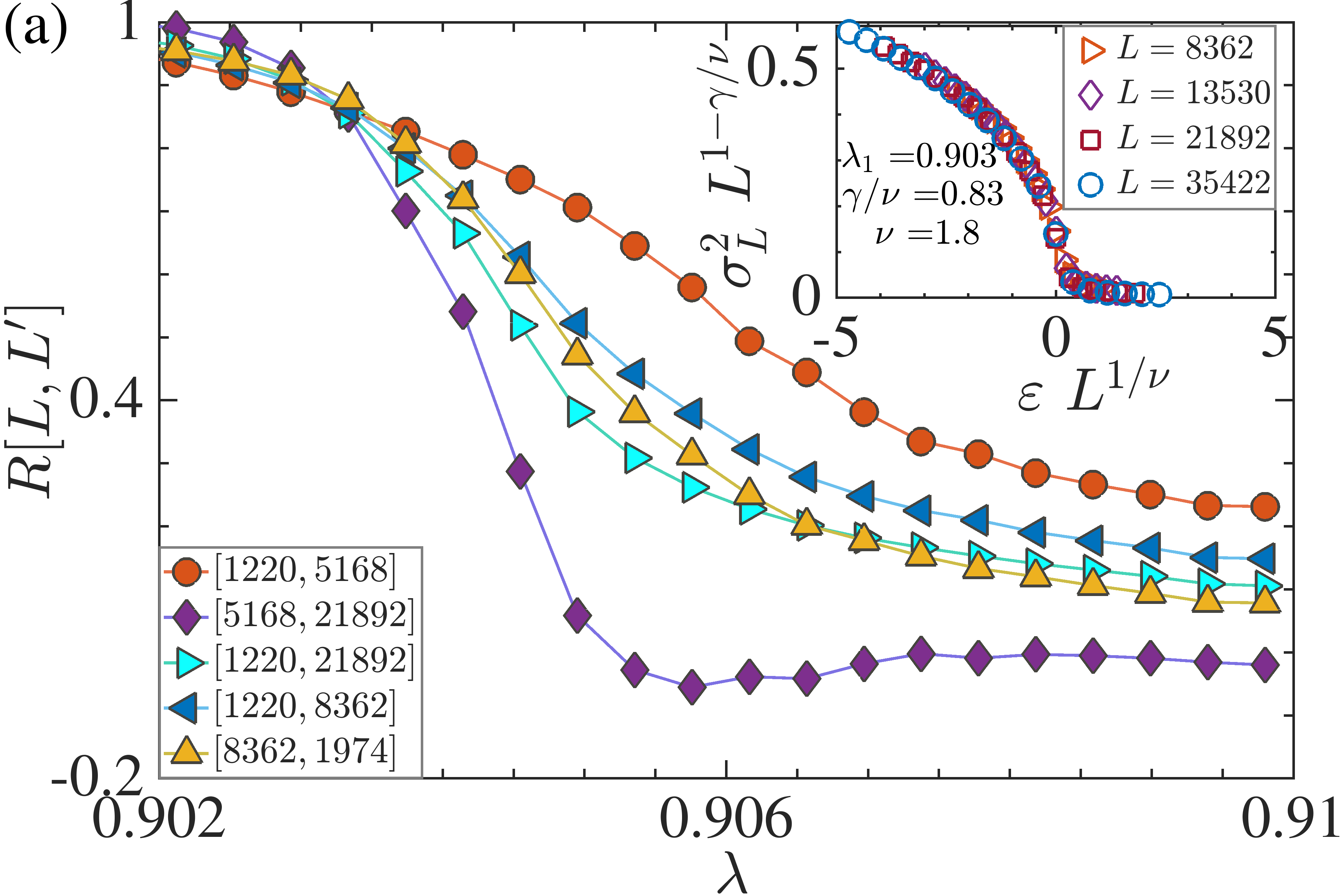}}
{\includegraphics[width=\linewidth]{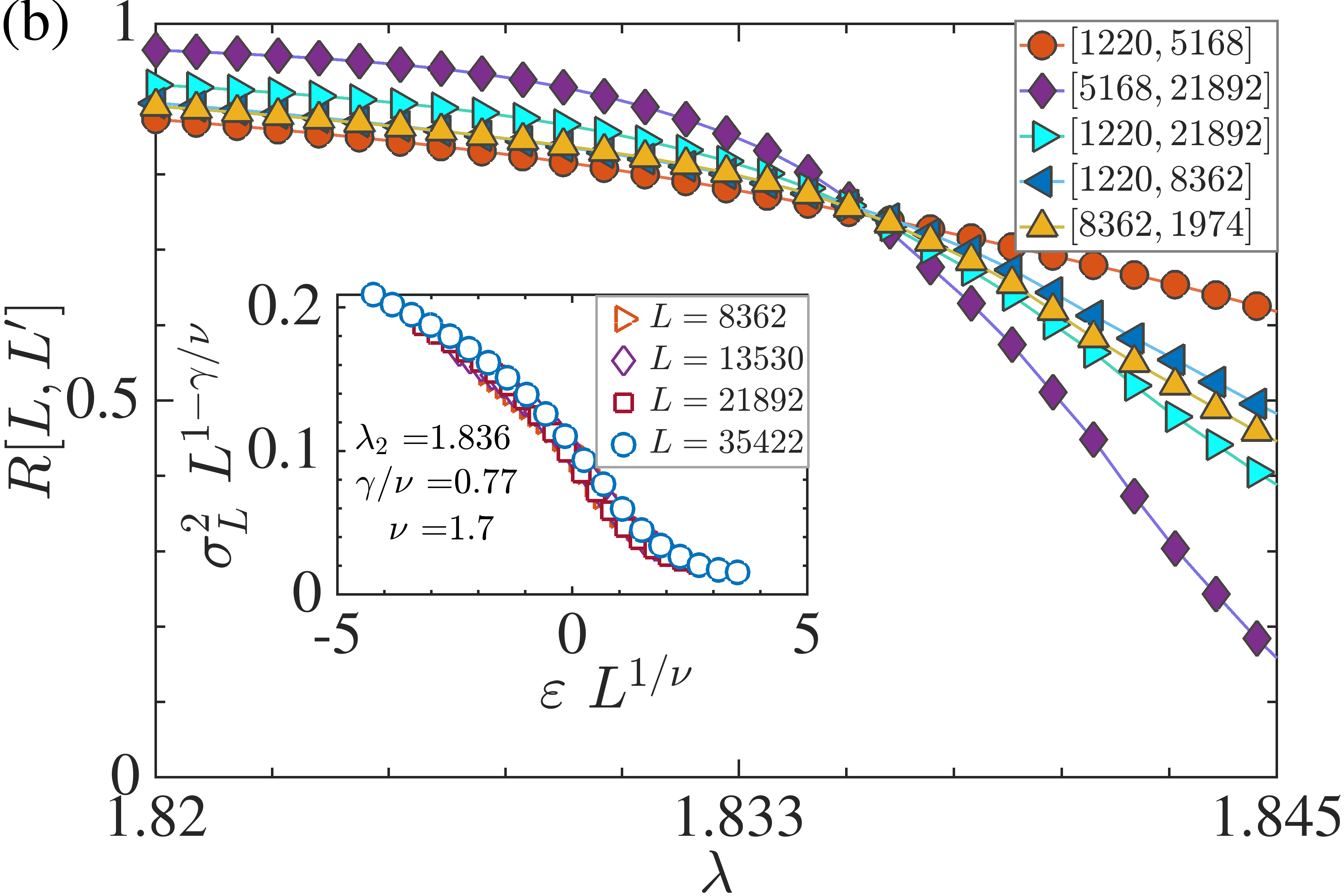}}
\caption{ Figure shows the plot of $R ~[L,L^{\prime}]$ in the vicinity of the first critical quasiperiodic potential strength $\lambda_{1}$ in (a) and for  the second critical quasiperiodic potential strength $\lambda_{2}$ in (b) corresponding to $\delta=2.2$. The insets show the data collapse with the $\sigma^{2}$ data in the vicinity of the first and the second critical points. Good data collapse is observed for both the transition points. The existence of single universal scaling functions can easily be inferred from the data collapse. We have done the calculations by taking an average over the states in the band with indices ($0-0.05$) for the first critical point and the states in the band  with indices ($0.45-0.5$) for the second critical point of the energy spectrum for the study.}
\label{fig:fig2}
\end{figure}

It is to be noted that the phase diagram shown in Fig.~\ref{fig:scheme1}(a) has been obtained by utilizing the behaviour of the average participation ratios, such as the inverse and the normalized participation ratios i.e. $\langle \rm{IPR}\rangle$ and the $\langle \rm{NPR}\rangle$ as a function of $\lambda$. Here, $\langle \cdot \rangle$ denotes the average taken over all the eigenstates corresponding to the Hamiltonian shown in Eq.~\ref{eqn:ham}~\cite{PhysRevLett.126.106803}. Before proceeding further, we first establish the transition points by analysing the behaviour of $\sigma$ which is directly related to the NPR of the states. From the definition, $\sigma$ for different lengths should approach zero at the localization transition. Therefore, it will be possible to estimate all the critical points by using the finite-size extrapolation of $\sigma$. For this purpose, we compute $\sigma$ by considering the eigenstates in a narrow band near the approximate transition boundaries. We plot $\sigma$ for different system sizes, namely, $L=3194,~5168,~8362,~13530$ and $21892$ as a function of $\lambda$ in Fig.~\ref{fig:sigma_lambda}(a-d) across the transition points $\lambda_1$, $\lambda_2$, $\lambda_3$ and $\lambda_4$ respectively. A finite size extrapolation reveals that for all the cases, $\sigma$ in the limit of $L\to\infty$ falls to a minimum after a critical $\lambda$  corresponding to different transitions. This defines the relevant range of $\lambda$ for our exploration of the critical properties. Once the limits of $\lambda$ around the critical transition points are identified we use them to calculate the function $R~[L,L^{\prime}]$ (see Eq.~(\ref{EqR})) as a function of $\lambda$. 

We first focus on the first localization transition which involves two critical points, such as $\lambda_1$ and $\lambda_2$ corresponding to the extended-intermediate and intermediate - localized phase transitions. Similar to the case of $\sigma$, for our analysis, to compute the function $R~[L,L^{\prime}]$ we use the eigenstates corresponding to a narrow band of the spectrum. We plot $R~[L,L^{\prime}]$ as a function of $\lambda$ for both the transitions around $\lambda_1$ and $\lambda_2$  in Figs.~\ref{fig:fig2}(a) and (b) respectively. The crossing of all the curves at a single point in both the figures (Fig.~\ref{fig:fig2}(a) and (b)) allows to obtain the critical points as $\lambda_1=0.903$ and $\lambda_2=1.836$. 
As already mentioned in Sec.~\ref{model}, following Eq.~\ref{gamma-1}, curves of $\sigma^2L^{1-\gamma/\nu}$ versus $\varepsilon L^{1/\nu}$ for different system sizes,  $L=8362,~13530,~21892$ and $35422$ collapse with the estimated critical strength $\lambda_1=0.903$. A perfect data collapse is obtained by considering $\gamma/\nu=0.83$ and $\nu=1.8$ for $\lambda_1=0.903$ as shown in the inset of Fig.~\ref{fig:fig2}(a). Similarly, for the second critical point ($\lambda_{2}=1.836$), a perfect data collapse is obtained by setting $\gamma/\nu=0.77$ and $\nu=1.7$ (inset of Fig.~\ref{fig:fig2}(b)). Note that the $\gamma/\nu$ considered for the data collapse matches fairly well with the ordinate corresponding to the points of intersection of $R~[L,L^{\prime}]$ as a function of $\lambda$ in Figs.~\ref{fig:fig2}(a) and (b).  
\begin{figure}[t]
\centering
{\includegraphics[width=\linewidth]{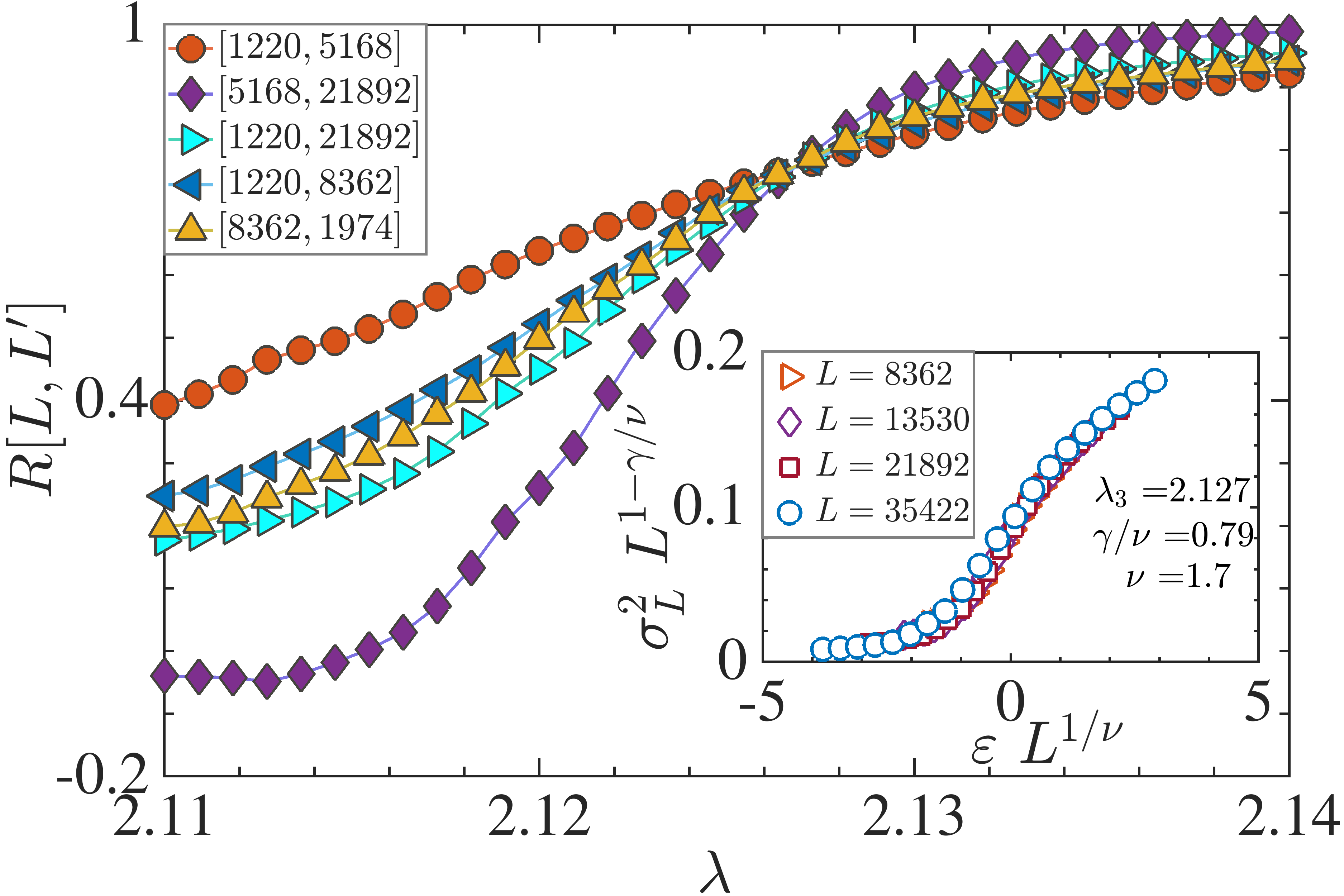}}
\caption{ The functions $R ~[L,L^{\prime}]$ are plotted as a function of $\lambda$  in the vicinity of the third critical point $\lambda_{3}$ at $\delta=2.2$. Inset shows the collapse of the $\sigma^{2}$ curves in the vicinity of the critical point. A data collapse is obtained by setting $\gamma/\nu=0.79$ and $\nu=1.7$ . We have done the calculations by taking an average over the states in the band  with indices ($0.45-0.5$) of the energy spectrum for the study.}
\label{fig:third_cp}
\end{figure}

We now turn our focus on to the second localization transition through the second critical region as depicted in Fig.~\ref{fig:scheme1}. This involves two transitions, namely, localized-intermediate and intermediate - localized transitions at the critical points $\lambda_3$ and $\lambda_4$ respectively. Following a similar scaling hypothesis as above, for the localized - intermediate transition, we obtain the crossing of $R~[L,L^{\prime}]$ data at a single point as depicted in Fig.~\ref{fig:third_cp} resulting in an accurate value of $\lambda_3=2.127$. Further, by using the values of  $\lambda_3$, a perfect data collapse is achieved in the  $\sigma^2L^{1-\gamma/\nu}$ versus $\varepsilon L^{1/\nu}$ plot by setting $\gamma/\nu=0.79$ and $\nu=1.7$  as shown in the inset of Fig.~\ref{fig:third_cp}. This suggests that the two transitions occurring at $\lambda_2$ and $\lambda_3$ corresponding to the transitions to and from the first localised phase that is intermediate - localized and localized - intermeidate phase transitions respectively belongs to the same universality class. 

\begin{figure}[!b]
\centering
{\includegraphics[width=\linewidth]{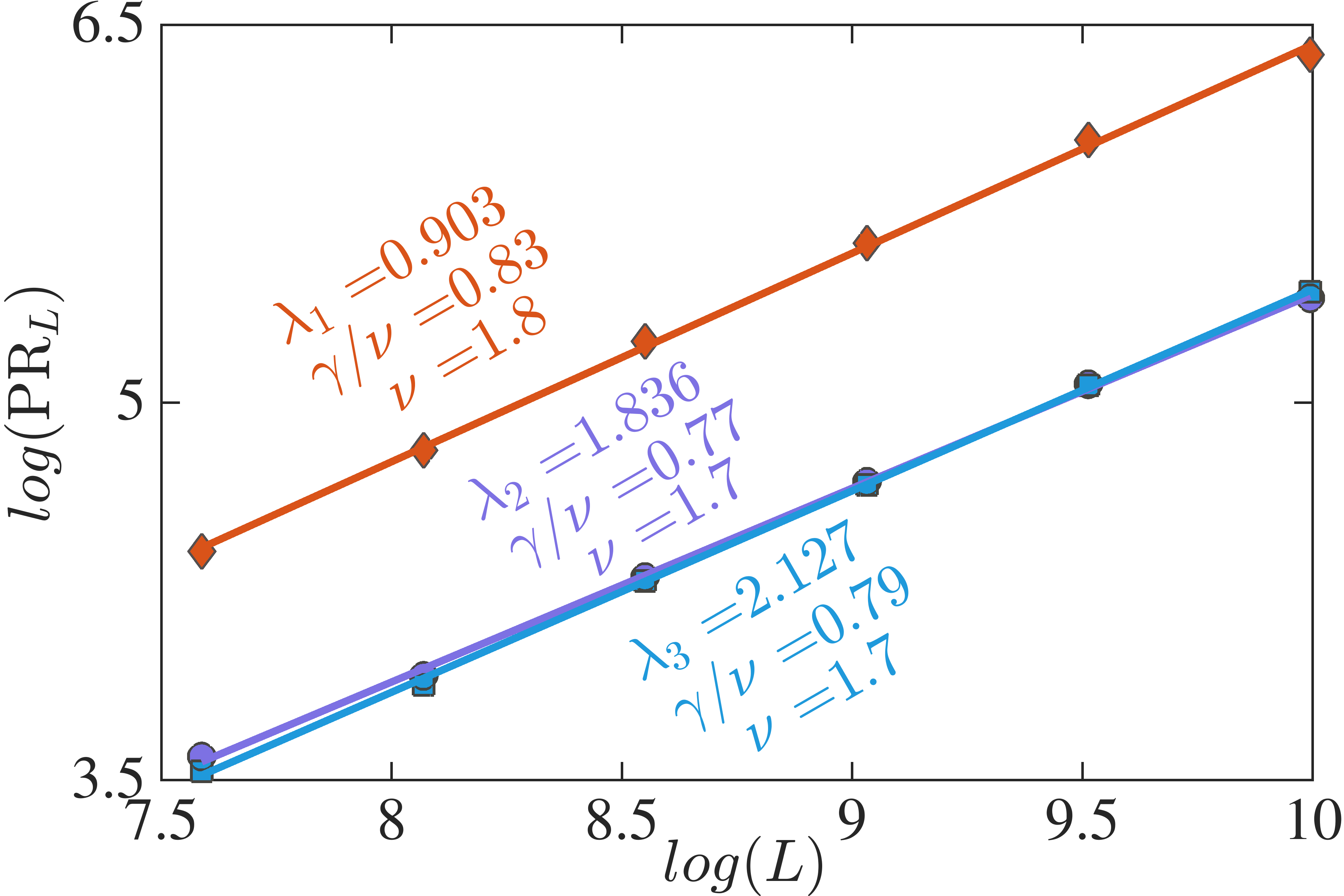}}
\caption{ The exponent ratio $\gamma/\nu$ is calculated via plotting the $log(PR_{L}) $ as a function of $log(L)$ for different system sizes of $L=1974,3194,5168,8362,13530,21892$ corresponding to three different critical points such as $\lambda_{1}$, $\lambda_{2}$, and $\lambda_{3}$.}
\label{fig:gamma_nu}
\end{figure}

It is now expected that the transition to the second localized phase i.e. the fourth transition at $\lambda_4$ falls under the same universality class as that of the second and third transitions at $\lambda_2$ and $\lambda_3$. However, in our scaling analysis we find an anomalous scaling behaviour of $R~[L,L^{\prime}]$ which is why we failed to achieve an accurate critical point $\lambda_4$ and the associated exponents. The actual reason for this behaviour can be attributed to the anomalous distribution of extended state (NPR$\neq0$) near the transition.
Before moving on to the multifractal analysis, we reconfirm the critical exponents from the scaling relation of the ${\rm{PR}}_L$ that denotes the participation ratio corresponding to different system sizes using Eq.~\ref{eq4} which can be written as ${\rm{PR}}_L \sim L^{\gamma/\nu}$. From this relation, a plot between $log({\rm{PR}}_L)$ and $log(L)$ for different lengths $L$ at the critical point should result in a straightline with slope $\gamma/\nu$. We performed this analysis at all the three critical points, such as $\lambda_1$, $\lambda_2$ and $\lambda_3$ in Fig.~\ref{fig:gamma_nu} and obtain the values of $\gamma/\nu$ as $0.83$, $0.77$ and $0.79$ respectively. 

\begin{figure}[!t]
\centering
{\includegraphics[width=\linewidth]{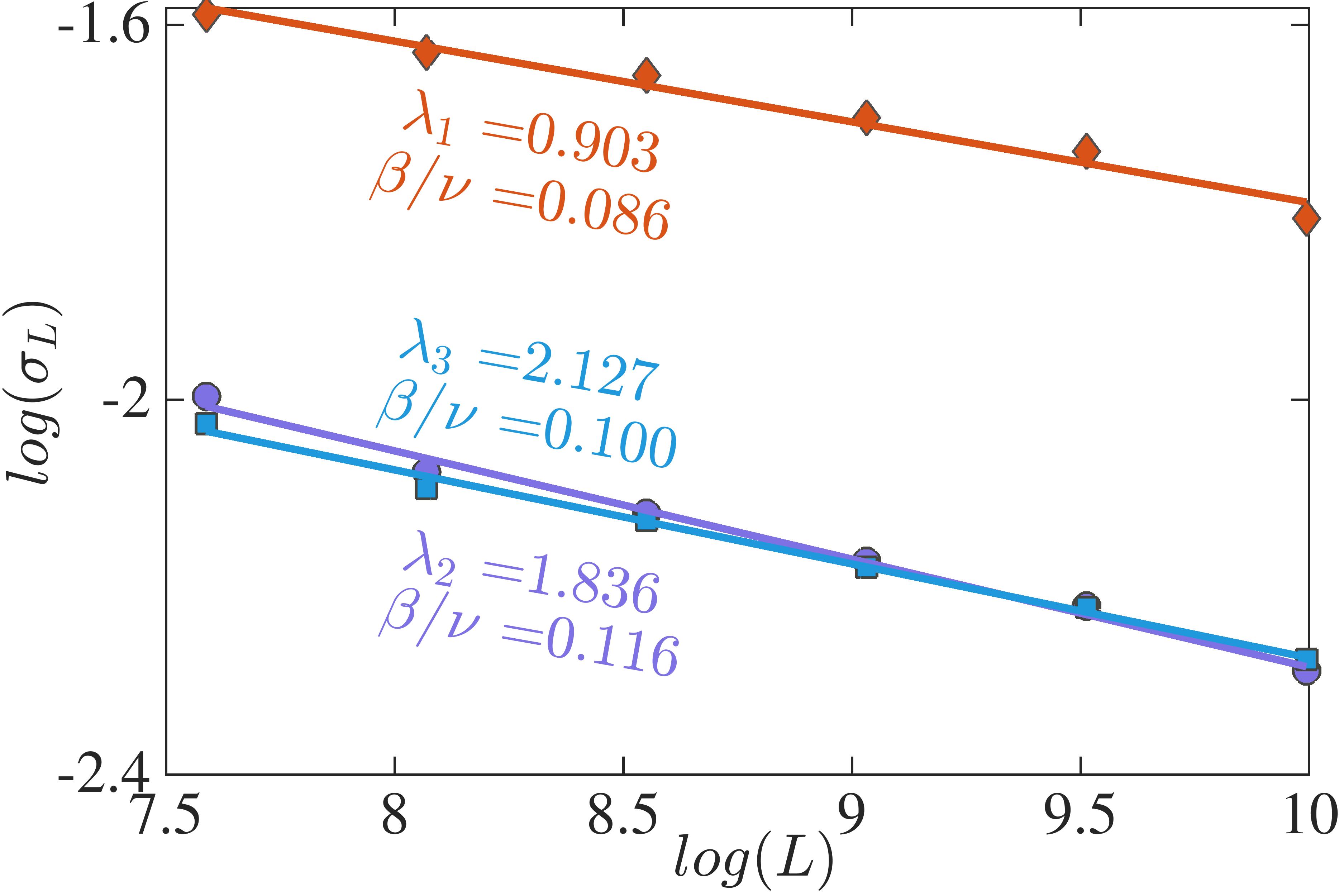}}
\caption{ The exponent ratio, $\beta/\nu$ is calculated via plotting the order parameter, $\sigma$ as a function of different system lengths ($L$) corresponding to three different critical potential strengths, namely, $\lambda_{1}$, $\lambda_{2}$, and $\lambda_{3}$.}
\label{fig:beta_nu}
\end{figure}

The exponents obtained in our analysis should satisfy a hyper-scaling law expressed as, \cite{Hashimoto_1992}
\begin{eqnarray}
\frac{2\beta}{\nu}+\frac{\gamma}{\nu}=1
\label{hsl1}
\end{eqnarray}
Using the hyper-scaling relation given in Eq.~\ref{hsl1}, it will be possible to extract another ratio of the exponents i.e. $\beta/\nu$ via 
\begin{equation}
 \frac{\beta}{\nu}=\frac{1}{2}\left(1-\frac{\gamma}{\nu}\right)
\end{equation}
Since at the critical point $\xi= L$, from Eq.~\ref{eq4} we have ${\sigma}\sim L^{-\beta/\nu}$. In order to establish the hyper-scaling relation we plot log($\sigma_{L}$) as a function of $log(L)$ for different system sizes corresponding to the three critical points $\lambda_{1}$, $\lambda_{2}$, and $\lambda_{3}$ mentioned in Fig.~\ref{fig:beta_nu}. The slopes of the curves  yield the exponent ratios $\beta/\nu=0.086,~0.116$ and $0.1$ for $\lambda_1=0.903$, $\lambda_2=1.836$ and $\lambda_3=2.127$ respectively. These values of the exponent ratios $\gamma/\nu$ ($0.83$, $0.77$ and $0.79$) and $\beta/\nu$ ($0.086$, $0.116$ and $0.100$) clearly satisfy the hyper-scaling relation (Eq.~\ref{hsl1}) at the critical points.

\subsection{Multifractal analysis} 
As already discussed the two localization transitions in this case occur through two intermediate regions. In analogy with the DL transition in the simple AA model and other models where the DL transition occurs through an intermediate region, we expect the eigenstates in the intermediate phases to be multifractal in nature. Thus, to explore deeper into the nature of the phases we perform a multifractal analysis~\cite{10.21468/SciPostPhys.4.5.025, deng2019one} of the eigenstates and calculate the associated fractal dimensions to arrive at an intuitive picture for the critical regions. In addition to that, we also study the energy spectrum corresponding to the critical regime via the scaling approach, hence calculate the Hausdorff dimension \cite{yao2019critical} of the energy spectrum. Note that for the multifractal analysis we consider the periodic boundary condition to avoid the effects arising from the edge states. 

A multifractal nature of the eigenstates can be identified via the generalized IPR and its scaling exponent $\tau_{q}$ \cite{10.21468/SciPostPhys.4.5.025, deng2019one,RevModPhys.80.1355} using the relation 
\begin{equation}
{\rm{IPR}}_{q}^{n}=\sum_{i=1}^{L} |\phi^{i}_{n}|^{2q} \to L^{-\tau_{q}},
\end{equation}
where $\tau_{q}$ is also known as the mass exponent and $q$ is a real number. The mass exponent vanishes for the localized states, whereas it varies linearly with the system dimension $d$ for the delocalized state as $\tau_{q}=d(q-1)$. 
Furthermore, the scaling exponents of the multifractal states can be characterized by a non-linear relation where $d$ (see above) is no longer an integer and further acquire a $q$ dependence which can be written as,
\begin{equation}
\tau_{q}=D_{q}(q-1),
\label{multifrac}
\end{equation}
where $D_{q}$ denotes the fractal dimension of the eigenstates.
Therefore, an extended and a localized state have respectively $1$ and $0$ as their fractal dimensions while an intermediate  value  of $D_{q}$ (between $1$ and $0$) denotes the fractal nature of the eigenstates. A spectrum possessing different fractal dimensions implies a multifractal behaviour of the eigenstates of the system. 
\begin{figure}[t]
\centering
{\includegraphics[width=\linewidth]{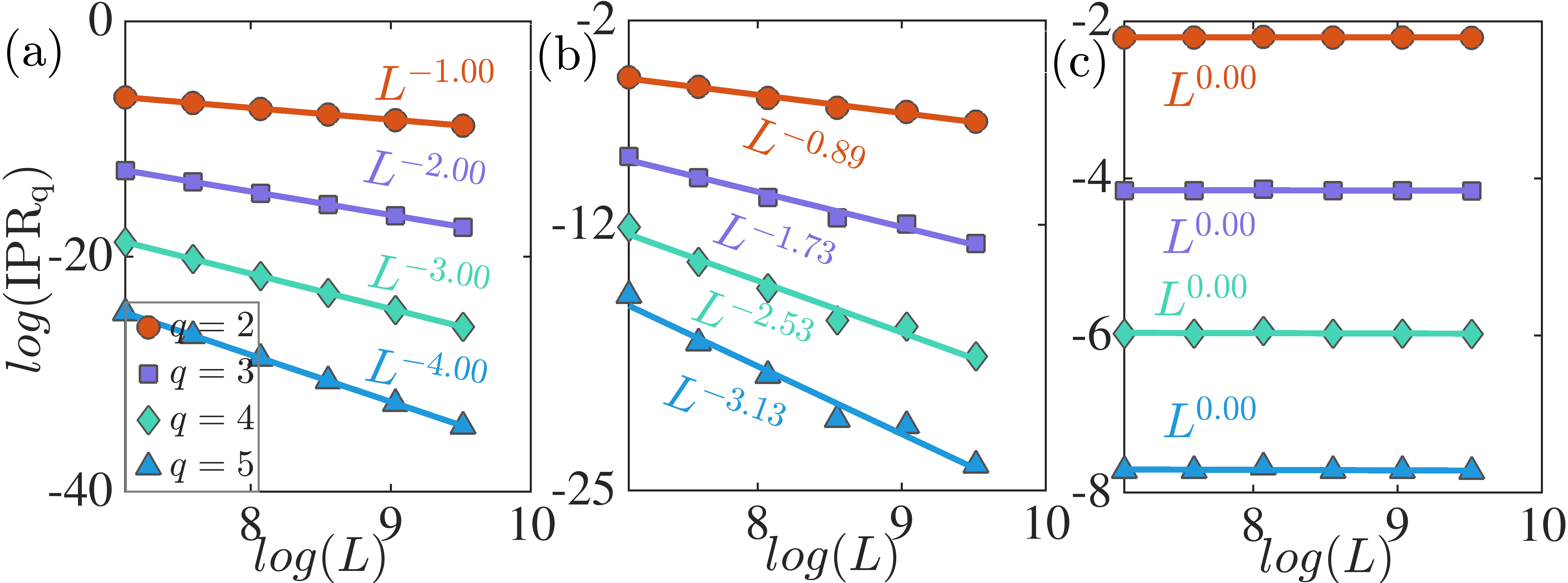}}
\caption{The generalized IPR is plotted as a function of different system sizes $L$ and corresponding to different moments of the intensity $q$. The slope of the curves are characterized by the mass exponent $\tau$. We have shown three distinguishing behavior of $\tau$ by considering the potential strength $\lambda$  in the extended, multifractal and localized regions in (a), (b) and (c) respectively. We have considered $\lambda=0.5$ and eigenstate index=$0.5$ in  (a) , $\lambda=0.903$ (first critical point) and eigenstate index= $0.1$ in (b), and  $\lambda=4$ and eigenstate index=$0.5$ in (c). For all the cases, we have taken $\delta=2.2$. }
\label{fig:tau_cal}
\end{figure}

\begin{figure}[b]
\centering
{\includegraphics[width=\linewidth]{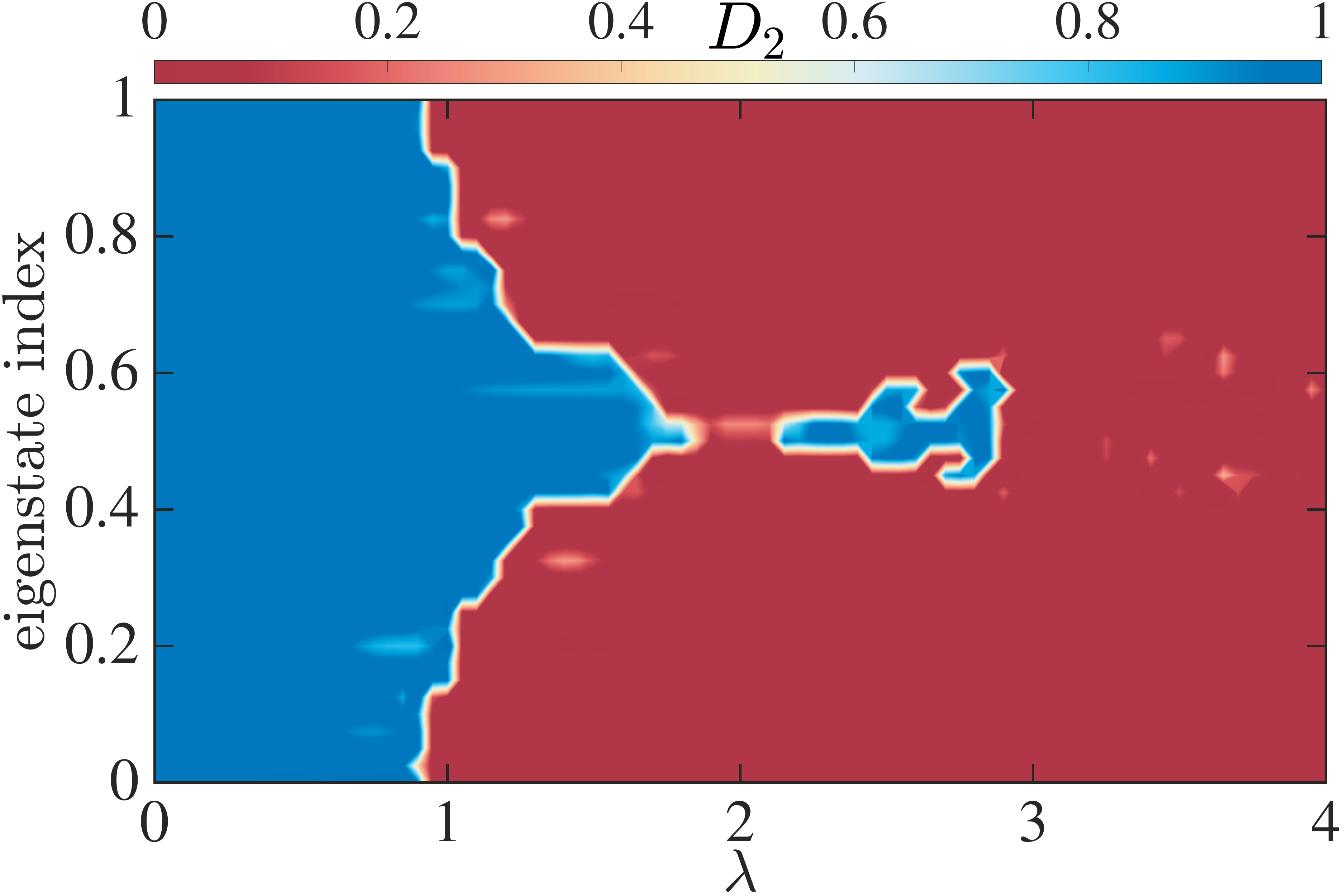}}
\caption{ The values of $D_{2}$ as a function of $\lambda$ and eigenstate index are plotted for $\delta=2.2$.}
\label{fig:tau_phase_diag}
\end{figure}
\begin{figure}[t]
\centering
{\includegraphics[width=\linewidth]{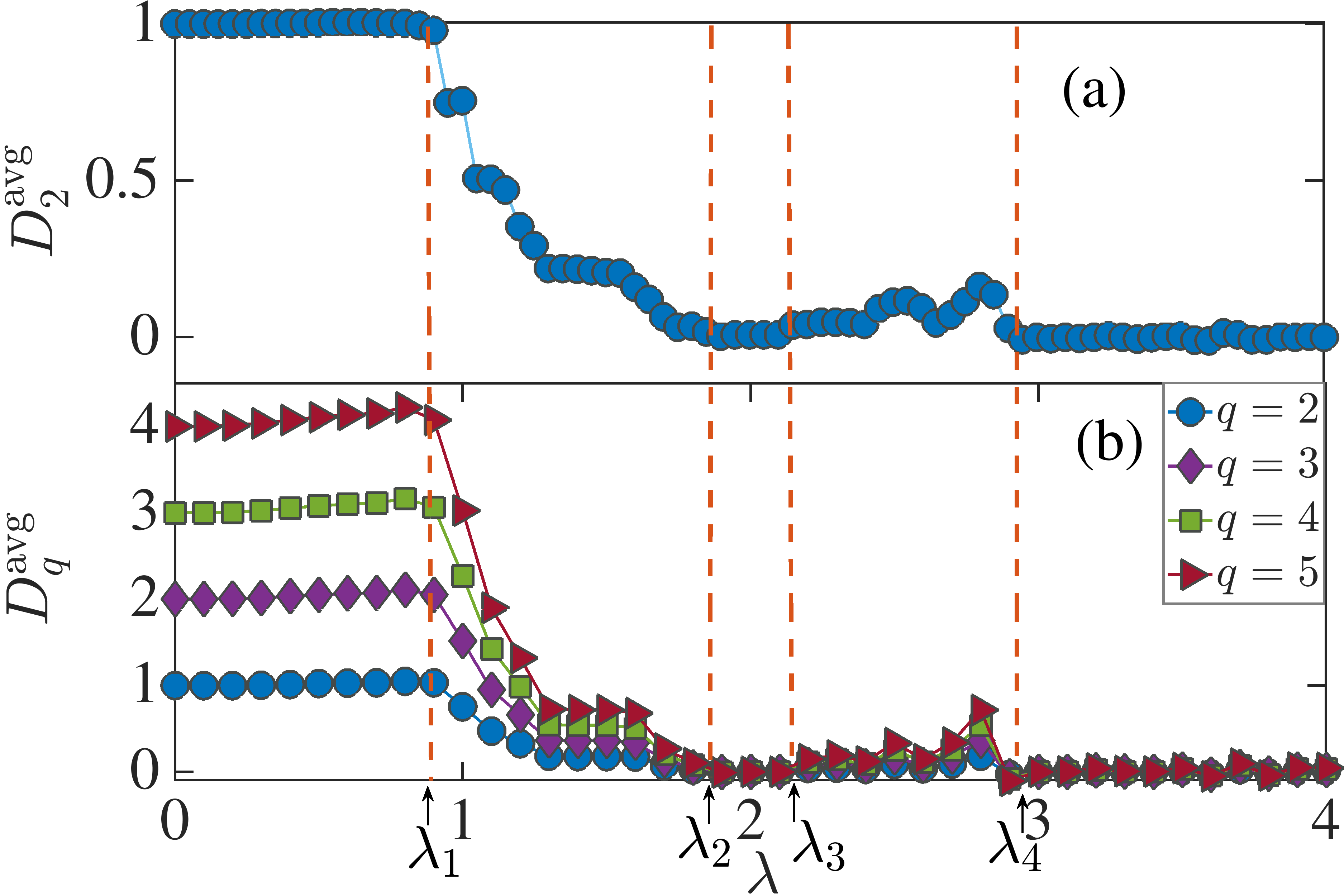}}
\caption{ $D_{2}^{avg}$  and $D_{q}^{avg}$ are plotted as a function of $\lambda$ in (a) and (b) respectively for $\delta=2.2$.  }
\label{fig:tau_D2}
\end{figure}
In our analysis, we first obtain the correlation dimension, denoted by $D_2$ corresponding to $q=2$ from the relation,
\begin{equation}
 {\rm{IPR}}_2^n \propto L^{-D_2}.
\end{equation}
$D_2$ can be obtained as the slope of the $log(\rm{IPR}_2)$ versus $log(L)$ plot corresponding to different states as shown in Fig.~\ref{fig:tau_cal}(red circles). Furthermore, in order to gain insights about the variation of $D_2$ over the entire spectrum, we plot $D_2$ as a function of eigenstate index and $\lambda$ at $\delta=2.2$ in Fig.~\ref{fig:tau_phase_diag} which clearly shows the existence of the extended, the localized and the multifractal states. Although, the expected re-entrant phase transitions can be seen from Fig.~\ref{fig:tau_phase_diag}, a clear understanding of this feature can be obtained from the average of $D_2$ over the eigenstates. In Fig.~\ref{fig:tau_D2}(a) we plot $D_2^{avg}$ as a function of $\lambda$, where the extended and localized phases are characterized by $D_2^{avg}=1$ and $0$ respectively. Whereas, $0 < D_2^{avg} < 1$ implies the presence of the states which are multifractal in nature. 

In addition to that, we also examine the variation of the exponents by considering  different values of $q>2$ that corresponding to the higher moments of the eigenstates. We obtain signatures which exactly match with the nature corresponding to the extended, the multifractal, and the localized state of the spectrum as shown in Fig.~\ref{fig:tau_cal}(a-c) respectively. A clear understanding of these features can also be achieved by plotting $D_q^{avg}$ for the entire range of $\lambda$. In Fig.~\ref{fig:tau_D2}(b) we plot $D_q^{avg}$ as a function of $\lambda$ for $q=2,~3,~4$ and $5$. The $q$ dependence of $D_q^{avg}$ indicate the presence of multifractal states. In Fig.~\ref{fig:tau_D2}, the different phase transitions are marked by the vertical dashed lines.

\begin{figure}[!b]
\centering
{\includegraphics[width=\linewidth]{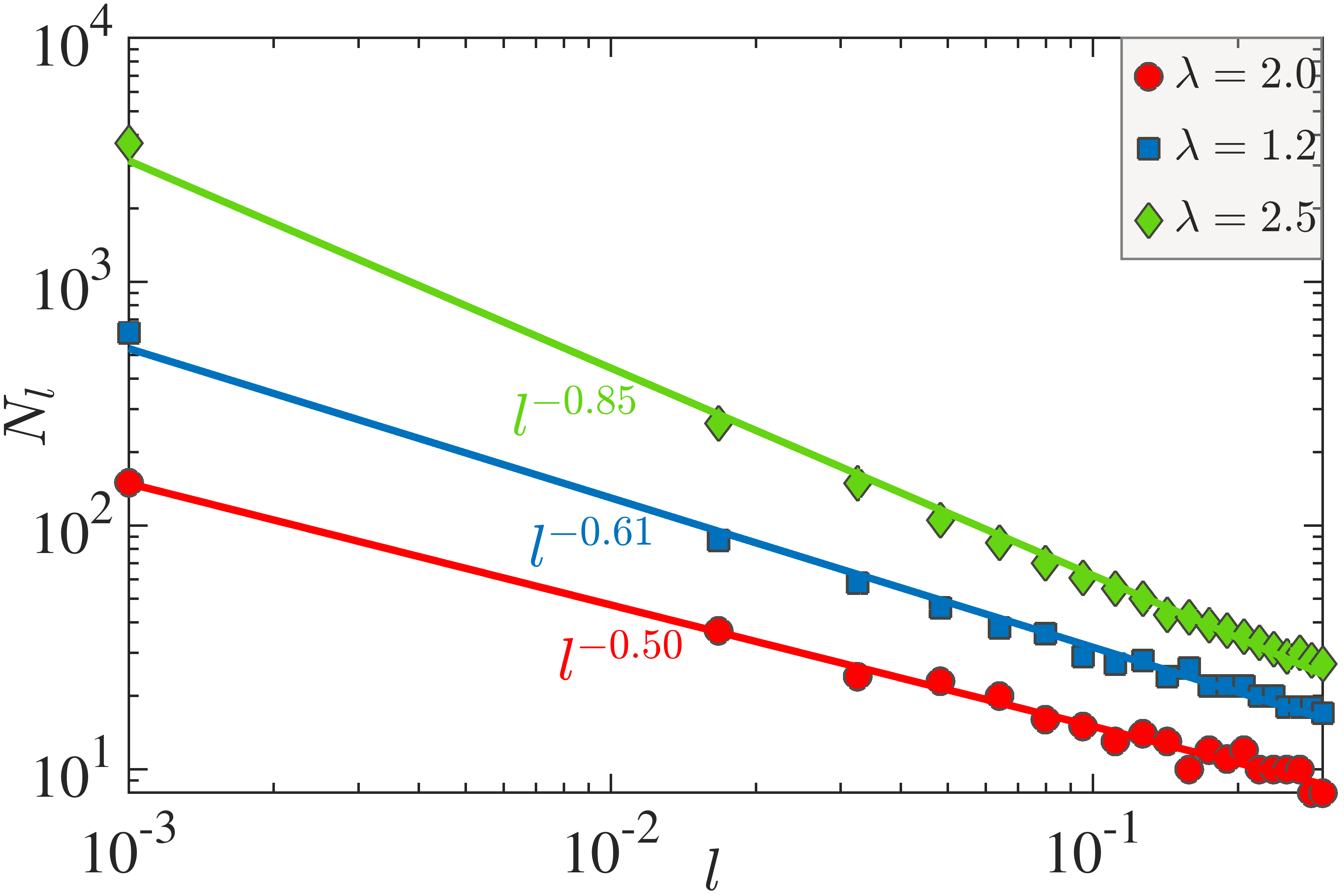}}
\caption{ Figure shows  $(N_{l}$ as a function of box length $l$ in the log-log scale corresponding to $\lambda=1.2$ ( blue squares) and $\lambda=2.5$ (green diamonds). For all the cases we have choosen $\delta=2.2$. For comparison, we have shown the result for the pure AA limit (red circles). The slopes of these plots give the Hausdorff dimensions which are obtained as $D_{H}=0.61$ and $0.85$ for $\lambda=1.2$ and $2.5$ respectively. Note that for the AA model $D_{H}=0.5$. The system size considered for the calculation is $L=13530$.}
\label{fig:Hausdorff_dim}
\end{figure}

\subsection{Hausdorff dimension}
The understanding of the details of the energy spectrum at the critical regime  can be complemented by computing the Hausdorff dimension of the system. A direct box-counting method is applied for this analysis~\cite{yao2019critical}. Considering the total number of boxes required is $N_{l}$ for a given box length  $l$ such that $N_{l}$ spans over the entire energy spectrum, $N_{l}$ shows a power-law behaviour with $l$ as,
\begin{equation}
N_{l} \propto l^{-D_{H}},
\label{Hausdorff}
\end{equation}
where, $D_{H}$ denotes the Hausdorff dimension corresponding to the energy spectrum. In our case, we compute the  $D_{H}$ by following Eq.~\ref{Hausdorff} in two different critical regions corresponding to $\lambda=1.2$ and $2.5$ which respectively denote the first and the second intermediate regimes. In Fig.~\ref{fig:Hausdorff_dim}, we plot $N_l$ as a function of $l$ which exhibits power law behaviour with exponent $D_H=0.61$ (blue squares)  and $0.85$ (green diamonds) for $\lambda=1.2$ and $2.5$ respectively. 
For comparison, we have plotted the corresponding AA limit ($\delta=1$, $\lambda_{A}=\lambda_{B}=\lambda=2$) (red circles)  which yields $D_H=0.5$~\cite{ikezawa1994energy}. From the analysis it is realized that the Hausdorff dimension in this case is different from the standard AA model.

\section{Conclusion} 
A one-dimensional quasiperiodic lattice model in the presence of hopping dimerization and a staggered on-site quasiperiodic potential exhibits re-entrant localization transitions. The transitions occur for a  range of dimerization strength through two intermediate phases resulting in four critical points. In this work, we characterize these transition points by using appropriate finite-size scaling laws for different order parameters. We also obtain the associate critical exponents which are found to obey the hyper-scaling laws. It is also observed that the second (intermediate - localized) and the third (localized - intermediate) phase transitions belong to the same universality class. Note that while we are able to accurately determine the first three critical points associated to the first localization transition, we fail to determine the last critical point of transition to the second localized phase. In addition to this, we have performed the multifractal analysis of the eigenstates and found that the states within the intermediate phases are multifractal in nature. Finally, we have calculated the Hausdorff dimension at the two critical regions which are found to be different from the standard AA limit.

\twocolumngrid

\bibliography{Critical_studies_SSH_AA}

\begin{thebibliography}{48}%
\makeatletter
\providecommand \@ifxundefined [1]{%
 \@ifx{#1\undefined}
}%
\providecommand \@ifnum [1]{%
 \ifnum #1\expandafter \@firstoftwo
 \else \expandafter \@secondoftwo
 \fi
}%
\providecommand \@ifx [1]{%
 \ifx #1\expandafter \@firstoftwo
 \else \expandafter \@secondoftwo
 \fi
}%
\providecommand \natexlab [1]{#1}%
\providecommand \enquote  [1]{``#1''}%
\providecommand \bibnamefont  [1]{#1}%
\providecommand \bibfnamefont [1]{#1}%
\providecommand \citenamefont [1]{#1}%
\providecommand \href@noop [0]{\@secondoftwo}%
\providecommand \href [0]{\begingroup \@sanitize@url \@href}%
\providecommand \@href[1]{\@@startlink{#1}\@@href}%
\providecommand \@@href[1]{\endgroup#1\@@endlink}%
\providecommand \@sanitize@url [0]{\catcode `\\12\catcode `\$12\catcode
  `\&12\catcode `\#12\catcode `\^12\catcode `\_12\catcode `\%12\relax}%
\providecommand \@@startlink[1]{}%
\providecommand \@@endlink[0]{}%
\providecommand \url  [0]{\begingroup\@sanitize@url \@url }%
\providecommand \@url [1]{\endgroup\@href {#1}{\urlprefix }}%
\providecommand \urlprefix  [0]{URL }%
\providecommand \Eprint [0]{\href }%
\providecommand \doibase [0]{http://dx.doi.org/}%
\providecommand \selectlanguage [0]{\@gobble}%
\providecommand \bibinfo  [0]{\@secondoftwo}%
\providecommand \bibfield  [0]{\@secondoftwo}%
\providecommand \translation [1]{[#1]}%
\providecommand \BibitemOpen [0]{}%
\providecommand \bibitemStop [0]{}%
\providecommand \bibitemNoStop [0]{.\EOS\space}%
\providecommand \EOS [0]{\spacefactor3000\relax}%
\providecommand \BibitemShut  [1]{\csname bibitem#1\endcsname}%
\let\auto@bib@innerbib\@empty
\bibitem [{\citenamefont {Roy}\ \emph {et~al.}(2021{\natexlab{a}})\citenamefont
  {Roy}, \citenamefont {Mishra}, \citenamefont {Tanatar},\ and\ \citenamefont
  {Basu}}]{PhysRevLett.126.106803}%
  \BibitemOpen
  \bibfield  {author} {\bibinfo {author} {\bibfnamefont {S.}~\bibnamefont
  {Roy}}, \bibinfo {author} {\bibfnamefont {T.}~\bibnamefont {Mishra}},
  \bibinfo {author} {\bibfnamefont {B.}~\bibnamefont {Tanatar}}, \ and\
  \bibinfo {author} {\bibfnamefont {S.}~\bibnamefont {Basu}},\ }\href {\doibase
  10.1103/PhysRevLett.126.106803} {\bibfield  {journal} {\bibinfo  {journal}
  {Phys. Rev. Lett.}\ }\textbf {\bibinfo {volume} {126}},\ \bibinfo {pages}
  {106803} (\bibinfo {year} {2021}{\natexlab{a}})}\BibitemShut {NoStop}%
\bibitem [{\citenamefont {Anderson}(1958)}]{PhysRev.109.1492}%
  \BibitemOpen
  \bibfield  {author} {\bibinfo {author} {\bibfnamefont {P.~W.}\ \bibnamefont
  {Anderson}},\ }\href {\doibase 10.1103/PhysRev.109.1492} {\bibfield
  {journal} {\bibinfo  {journal} {Phys. Rev.}\ }\textbf {\bibinfo {volume}
  {109}},\ \bibinfo {pages} {1492} (\bibinfo {year} {1958})}\BibitemShut
  {NoStop}%
\bibitem [{\citenamefont {Abrahams}\ \emph {et~al.}(1979)\citenamefont
  {Abrahams}, \citenamefont {Anderson}, \citenamefont {Licciardello},\ and\
  \citenamefont {Ramakrishnan}}]{PhysRevLett.42.673}%
  \BibitemOpen
  \bibfield  {author} {\bibinfo {author} {\bibfnamefont {E.}~\bibnamefont
  {Abrahams}}, \bibinfo {author} {\bibfnamefont {P.~W.}\ \bibnamefont
  {Anderson}}, \bibinfo {author} {\bibfnamefont {D.~C.}\ \bibnamefont
  {Licciardello}}, \ and\ \bibinfo {author} {\bibfnamefont {T.~V.}\
  \bibnamefont {Ramakrishnan}},\ }\href {\doibase 10.1103/PhysRevLett.42.673}
  {\bibfield  {journal} {\bibinfo  {journal} {Phys. Rev. Lett.}\ }\textbf
  {\bibinfo {volume} {42}},\ \bibinfo {pages} {673} (\bibinfo {year}
  {1979})}\BibitemShut {NoStop}%
\bibitem [{\citenamefont {Sokoloff}(1985)}]{SOKOLOFF1985189}%
  \BibitemOpen
  \bibfield  {author} {\bibinfo {author} {\bibfnamefont {J.}~\bibnamefont
  {Sokoloff}},\ }\href {\doibase https://doi.org/10.1016/0370-1573(85)90088-2}
  {\bibfield  {journal} {\bibinfo  {journal} {Physics Reports}\ }\textbf
  {\bibinfo {volume} {126}},\ \bibinfo {pages} {189} (\bibinfo {year}
  {1985})}\BibitemShut {NoStop}%
\bibitem [{\citenamefont {Grempel}\ \emph {et~al.}(1982)\citenamefont
  {Grempel}, \citenamefont {Fishman},\ and\ \citenamefont
  {Prange}}]{PhysRevLett.49.833}%
  \BibitemOpen
  \bibfield  {author} {\bibinfo {author} {\bibfnamefont {D.~R.}\ \bibnamefont
  {Grempel}}, \bibinfo {author} {\bibfnamefont {S.}~\bibnamefont {Fishman}}, \
  and\ \bibinfo {author} {\bibfnamefont {R.~E.}\ \bibnamefont {Prange}},\
  }\href {\doibase 10.1103/PhysRevLett.49.833} {\bibfield  {journal} {\bibinfo
  {journal} {Phys. Rev. Lett.}\ }\textbf {\bibinfo {volume} {49}},\ \bibinfo
  {pages} {833} (\bibinfo {year} {1982})}\BibitemShut {NoStop}%
\bibitem [{\citenamefont {Simon}(1985)}]{simon1985almost}%
  \BibitemOpen
  \bibfield  {author} {\bibinfo {author} {\bibfnamefont {B.}~\bibnamefont
  {Simon}},\ }\href@noop {} {\bibfield  {journal} {\bibinfo  {journal} {Annals
  of Physics}\ }\textbf {\bibinfo {volume} {159}},\ \bibinfo {pages} {157}
  (\bibinfo {year} {1985})}\BibitemShut {NoStop}%
\bibitem [{\citenamefont {Kohmoto}(1983{\natexlab{a}})}]{PhysRevLett.51.1198}%
  \BibitemOpen
  \bibfield  {author} {\bibinfo {author} {\bibfnamefont {M.}~\bibnamefont
  {Kohmoto}},\ }\href {\doibase 10.1103/PhysRevLett.51.1198} {\bibfield
  {journal} {\bibinfo  {journal} {Phys. Rev. Lett.}\ }\textbf {\bibinfo
  {volume} {51}},\ \bibinfo {pages} {1198} (\bibinfo {year}
  {1983}{\natexlab{a}})}\BibitemShut {NoStop}%
\bibitem [{\citenamefont {Ostlund}\ \emph {et~al.}(1983)\citenamefont
  {Ostlund}, \citenamefont {Pandit}, \citenamefont {Rand}, \citenamefont
  {Schellnhuber},\ and\ \citenamefont {Siggia}}]{PhysRevLett.50.1873}%
  \BibitemOpen
  \bibfield  {author} {\bibinfo {author} {\bibfnamefont {S.}~\bibnamefont
  {Ostlund}}, \bibinfo {author} {\bibfnamefont {R.}~\bibnamefont {Pandit}},
  \bibinfo {author} {\bibfnamefont {D.}~\bibnamefont {Rand}}, \bibinfo {author}
  {\bibfnamefont {H.~J.}\ \bibnamefont {Schellnhuber}}, \ and\ \bibinfo
  {author} {\bibfnamefont {E.~D.}\ \bibnamefont {Siggia}},\ }\href {\doibase
  10.1103/PhysRevLett.50.1873} {\bibfield  {journal} {\bibinfo  {journal}
  {Phys. Rev. Lett.}\ }\textbf {\bibinfo {volume} {50}},\ \bibinfo {pages}
  {1873} (\bibinfo {year} {1983})}\BibitemShut {NoStop}%
\bibitem [{\citenamefont {Aubry}\ and\ \citenamefont
  {Andr{\'e}}(1980)}]{aubry1980analyticity}%
  \BibitemOpen
  \bibfield  {author} {\bibinfo {author} {\bibfnamefont {S.}~\bibnamefont
  {Aubry}}\ and\ \bibinfo {author} {\bibfnamefont {G.}~\bibnamefont
  {Andr{\'e}}},\ }\href@noop {} {\bibfield  {journal} {\bibinfo  {journal}
  {Ann. Israel Phys. Soc}\ }\textbf {\bibinfo {volume} {3}},\ \bibinfo {pages}
  {18} (\bibinfo {year} {1980})}\BibitemShut {NoStop}%
\bibitem [{\citenamefont {Lahini}\ \emph {et~al.}(2009)\citenamefont {Lahini},
  \citenamefont {Pugatch}, \citenamefont {Pozzi}, \citenamefont {Sorel},
  \citenamefont {Morandotti}, \citenamefont {Davidson},\ and\ \citenamefont
  {Silberberg}}]{PhysRevLett.103.013901}%
  \BibitemOpen
  \bibfield  {author} {\bibinfo {author} {\bibfnamefont {Y.}~\bibnamefont
  {Lahini}}, \bibinfo {author} {\bibfnamefont {R.}~\bibnamefont {Pugatch}},
  \bibinfo {author} {\bibfnamefont {F.}~\bibnamefont {Pozzi}}, \bibinfo
  {author} {\bibfnamefont {M.}~\bibnamefont {Sorel}}, \bibinfo {author}
  {\bibfnamefont {R.}~\bibnamefont {Morandotti}}, \bibinfo {author}
  {\bibfnamefont {N.}~\bibnamefont {Davidson}}, \ and\ \bibinfo {author}
  {\bibfnamefont {Y.}~\bibnamefont {Silberberg}},\ }\href {\doibase
  10.1103/PhysRevLett.103.013901} {\bibfield  {journal} {\bibinfo  {journal}
  {Phys. Rev. Lett.}\ }\textbf {\bibinfo {volume} {103}},\ \bibinfo {pages}
  {013901} (\bibinfo {year} {2009})}\BibitemShut {NoStop}%
\bibitem [{\citenamefont {Merlin}\ \emph {et~al.}(1985)\citenamefont {Merlin},
  \citenamefont {Bajema}, \citenamefont {Clarke}, \citenamefont {Juang},\ and\
  \citenamefont {Bhattacharya}}]{PhysRevLett.55.1768}%
  \BibitemOpen
  \bibfield  {author} {\bibinfo {author} {\bibfnamefont {R.}~\bibnamefont
  {Merlin}}, \bibinfo {author} {\bibfnamefont {K.}~\bibnamefont {Bajema}},
  \bibinfo {author} {\bibfnamefont {R.}~\bibnamefont {Clarke}}, \bibinfo
  {author} {\bibfnamefont {F.~Y.}\ \bibnamefont {Juang}}, \ and\ \bibinfo
  {author} {\bibfnamefont {P.~K.}\ \bibnamefont {Bhattacharya}},\ }\href
  {\doibase 10.1103/PhysRevLett.55.1768} {\bibfield  {journal} {\bibinfo
  {journal} {Phys. Rev. Lett.}\ }\textbf {\bibinfo {volume} {55}},\ \bibinfo
  {pages} {1768} (\bibinfo {year} {1985})}\BibitemShut {NoStop}%
\bibitem [{\citenamefont {Tanese}\ \emph {et~al.}(2014)\citenamefont {Tanese},
  \citenamefont {Gurevich}, \citenamefont {Baboux}, \citenamefont {Jacqmin},
  \citenamefont {Lema\^{\i}tre}, \citenamefont {Galopin}, \citenamefont
  {Sagnes}, \citenamefont {Amo}, \citenamefont {Bloch},\ and\ \citenamefont
  {Akkermans}}]{PhysRevLett.112.146404}%
  \BibitemOpen
  \bibfield  {author} {\bibinfo {author} {\bibfnamefont {D.}~\bibnamefont
  {Tanese}}, \bibinfo {author} {\bibfnamefont {E.}~\bibnamefont {Gurevich}},
  \bibinfo {author} {\bibfnamefont {F.}~\bibnamefont {Baboux}}, \bibinfo
  {author} {\bibfnamefont {T.}~\bibnamefont {Jacqmin}}, \bibinfo {author}
  {\bibfnamefont {A.}~\bibnamefont {Lema\^{\i}tre}}, \bibinfo {author}
  {\bibfnamefont {E.}~\bibnamefont {Galopin}}, \bibinfo {author} {\bibfnamefont
  {I.}~\bibnamefont {Sagnes}}, \bibinfo {author} {\bibfnamefont
  {A.}~\bibnamefont {Amo}}, \bibinfo {author} {\bibfnamefont {J.}~\bibnamefont
  {Bloch}}, \ and\ \bibinfo {author} {\bibfnamefont {E.}~\bibnamefont
  {Akkermans}},\ }\href {\doibase 10.1103/PhysRevLett.112.146404} {\bibfield
  {journal} {\bibinfo  {journal} {Phys. Rev. Lett.}\ }\textbf {\bibinfo
  {volume} {112}},\ \bibinfo {pages} {146404} (\bibinfo {year}
  {2014})}\BibitemShut {NoStop}%
\bibitem [{\citenamefont {Roushan}\ \emph {et~al.}(2017)\citenamefont
  {Roushan}, \citenamefont {Neill}, \citenamefont {Tangpanitanon},
  \citenamefont {Bastidas}, \citenamefont {Megrant}, \citenamefont {Barends},
  \citenamefont {Chen}, \citenamefont {Chen}, \citenamefont {Chiaro},
  \citenamefont {Dunsworth} \emph {et~al.}}]{roushan2017spectroscopic}%
  \BibitemOpen
  \bibfield  {author} {\bibinfo {author} {\bibfnamefont {P.}~\bibnamefont
  {Roushan}}, \bibinfo {author} {\bibfnamefont {C.}~\bibnamefont {Neill}},
  \bibinfo {author} {\bibfnamefont {J.}~\bibnamefont {Tangpanitanon}}, \bibinfo
  {author} {\bibfnamefont {V.~M.}\ \bibnamefont {Bastidas}}, \bibinfo {author}
  {\bibfnamefont {A.}~\bibnamefont {Megrant}}, \bibinfo {author} {\bibfnamefont
  {R.}~\bibnamefont {Barends}}, \bibinfo {author} {\bibfnamefont
  {Y.}~\bibnamefont {Chen}}, \bibinfo {author} {\bibfnamefont {Z.}~\bibnamefont
  {Chen}}, \bibinfo {author} {\bibfnamefont {B.}~\bibnamefont {Chiaro}},
  \bibinfo {author} {\bibfnamefont {A.}~\bibnamefont {Dunsworth}},  \emph
  {et~al.},\ }\href@noop {} {\bibfield  {journal} {\bibinfo  {journal}
  {Science}\ }\textbf {\bibinfo {volume} {358}},\ \bibinfo {pages} {1175}
  (\bibinfo {year} {2017})}\BibitemShut {NoStop}%
\bibitem [{\citenamefont {Billy}\ \emph {et~al.}(2008)\citenamefont {Billy},
  \citenamefont {Josse}, \citenamefont {Zuo}, \citenamefont {Bernard},
  \citenamefont {Hambrecht}, \citenamefont {Lugan}, \citenamefont
  {Cl{\'e}ment}, \citenamefont {Sanchez-Palencia}, \citenamefont {Bouyer},\
  and\ \citenamefont {Aspect}}]{billy2008direct}%
  \BibitemOpen
  \bibfield  {author} {\bibinfo {author} {\bibfnamefont {J.}~\bibnamefont
  {Billy}}, \bibinfo {author} {\bibfnamefont {V.}~\bibnamefont {Josse}},
  \bibinfo {author} {\bibfnamefont {Z.}~\bibnamefont {Zuo}}, \bibinfo {author}
  {\bibfnamefont {A.}~\bibnamefont {Bernard}}, \bibinfo {author} {\bibfnamefont
  {B.}~\bibnamefont {Hambrecht}}, \bibinfo {author} {\bibfnamefont
  {P.}~\bibnamefont {Lugan}}, \bibinfo {author} {\bibfnamefont
  {D.}~\bibnamefont {Cl{\'e}ment}}, \bibinfo {author} {\bibfnamefont
  {L.}~\bibnamefont {Sanchez-Palencia}}, \bibinfo {author} {\bibfnamefont
  {P.}~\bibnamefont {Bouyer}}, \ and\ \bibinfo {author} {\bibfnamefont
  {A.}~\bibnamefont {Aspect}},\ }\href@noop {} {\bibfield  {journal} {\bibinfo
  {journal} {Nature}\ }\textbf {\bibinfo {volume} {453}},\ \bibinfo {pages}
  {891} (\bibinfo {year} {2008})}\BibitemShut {NoStop}%
\bibitem [{\citenamefont {Aspect}\ and\ \citenamefont
  {Inguscio}(2009)}]{aspect2009anderson}%
  \BibitemOpen
  \bibfield  {author} {\bibinfo {author} {\bibfnamefont {A.}~\bibnamefont
  {Aspect}}\ and\ \bibinfo {author} {\bibfnamefont {M.}~\bibnamefont
  {Inguscio}},\ }\href@noop {} {\bibfield  {journal} {\bibinfo  {journal}
  {Phys. Today}\ }\textbf {\bibinfo {volume} {62}},\ \bibinfo {pages} {30}
  (\bibinfo {year} {2009})}\BibitemShut {NoStop}%
\bibitem [{\citenamefont {Lahini}\ \emph {et~al.}(2008)\citenamefont {Lahini},
  \citenamefont {Avidan}, \citenamefont {Pozzi}, \citenamefont {Sorel},
  \citenamefont {Morandotti}, \citenamefont {Christodoulides},\ and\
  \citenamefont {Silberberg}}]{lahini2008anderson}%
  \BibitemOpen
  \bibfield  {author} {\bibinfo {author} {\bibfnamefont {Y.}~\bibnamefont
  {Lahini}}, \bibinfo {author} {\bibfnamefont {A.}~\bibnamefont {Avidan}},
  \bibinfo {author} {\bibfnamefont {F.}~\bibnamefont {Pozzi}}, \bibinfo
  {author} {\bibfnamefont {M.}~\bibnamefont {Sorel}}, \bibinfo {author}
  {\bibfnamefont {R.}~\bibnamefont {Morandotti}}, \bibinfo {author}
  {\bibfnamefont {D.~N.}\ \bibnamefont {Christodoulides}}, \ and\ \bibinfo
  {author} {\bibfnamefont {Y.}~\bibnamefont {Silberberg}},\ }\href@noop {}
  {\bibfield  {journal} {\bibinfo  {journal} {Phys. Rev. Lett.}\ }\textbf
  {\bibinfo {volume} {100}},\ \bibinfo {pages} {013906} (\bibinfo {year}
  {2008})}\BibitemShut {NoStop}%
\bibitem [{\citenamefont {L\"uschen}\ \emph {et~al.}(2018)\citenamefont
  {L\"uschen}, \citenamefont {Scherg}, \citenamefont {Kohlert}, \citenamefont
  {Schreiber}, \citenamefont {Bordia}, \citenamefont {Li}, \citenamefont
  {Das~Sarma},\ and\ \citenamefont {Bloch}}]{PhysRevLett.120.160404}%
  \BibitemOpen
  \bibfield  {author} {\bibinfo {author} {\bibfnamefont {H.~P.}\ \bibnamefont
  {L\"uschen}}, \bibinfo {author} {\bibfnamefont {S.}~\bibnamefont {Scherg}},
  \bibinfo {author} {\bibfnamefont {T.}~\bibnamefont {Kohlert}}, \bibinfo
  {author} {\bibfnamefont {M.}~\bibnamefont {Schreiber}}, \bibinfo {author}
  {\bibfnamefont {P.}~\bibnamefont {Bordia}}, \bibinfo {author} {\bibfnamefont
  {X.}~\bibnamefont {Li}}, \bibinfo {author} {\bibfnamefont {S.}~\bibnamefont
  {Das~Sarma}}, \ and\ \bibinfo {author} {\bibfnamefont {I.}~\bibnamefont
  {Bloch}},\ }\href {\doibase 10.1103/PhysRevLett.120.160404} {\bibfield
  {journal} {\bibinfo  {journal} {Phys. Rev. Lett.}\ }\textbf {\bibinfo
  {volume} {120}},\ \bibinfo {pages} {160404} (\bibinfo {year}
  {2018})}\BibitemShut {NoStop}%
\bibitem [{\citenamefont {Fallani}\ \emph {et~al.}(2007)\citenamefont
  {Fallani}, \citenamefont {Lye}, \citenamefont {Guarrera}, \citenamefont
  {Fort},\ and\ \citenamefont {Inguscio}}]{PhysRevLett.98.130404}%
  \BibitemOpen
  \bibfield  {author} {\bibinfo {author} {\bibfnamefont {L.}~\bibnamefont
  {Fallani}}, \bibinfo {author} {\bibfnamefont {J.~E.}\ \bibnamefont {Lye}},
  \bibinfo {author} {\bibfnamefont {V.}~\bibnamefont {Guarrera}}, \bibinfo
  {author} {\bibfnamefont {C.}~\bibnamefont {Fort}}, \ and\ \bibinfo {author}
  {\bibfnamefont {M.}~\bibnamefont {Inguscio}},\ }\href {\doibase
  10.1103/PhysRevLett.98.130404} {\bibfield  {journal} {\bibinfo  {journal}
  {Phys. Rev. Lett.}\ }\textbf {\bibinfo {volume} {98}},\ \bibinfo {pages}
  {130404} (\bibinfo {year} {2007})}\BibitemShut {NoStop}%
\bibitem [{\citenamefont {Sanchez-Palencia}\ and\ \citenamefont
  {Santos}(2005)}]{PhysRevA.72.053607}%
  \BibitemOpen
  \bibfield  {author} {\bibinfo {author} {\bibfnamefont {L.}~\bibnamefont
  {Sanchez-Palencia}}\ and\ \bibinfo {author} {\bibfnamefont {L.}~\bibnamefont
  {Santos}},\ }\href {\doibase 10.1103/PhysRevA.72.053607} {\bibfield
  {journal} {\bibinfo  {journal} {Phys. Rev. A}\ }\textbf {\bibinfo {volume}
  {72}},\ \bibinfo {pages} {053607} (\bibinfo {year} {2005})}\BibitemShut
  {NoStop}%
\bibitem [{\citenamefont {Viebahn}\ \emph {et~al.}(2019)\citenamefont
  {Viebahn}, \citenamefont {Sbroscia}, \citenamefont {Carter}, \citenamefont
  {Yu},\ and\ \citenamefont {Schneider}}]{viebahn2019matter}%
  \BibitemOpen
  \bibfield  {author} {\bibinfo {author} {\bibfnamefont {K.}~\bibnamefont
  {Viebahn}}, \bibinfo {author} {\bibfnamefont {M.}~\bibnamefont {Sbroscia}},
  \bibinfo {author} {\bibfnamefont {E.}~\bibnamefont {Carter}}, \bibinfo
  {author} {\bibfnamefont {J.-C.}\ \bibnamefont {Yu}}, \ and\ \bibinfo {author}
  {\bibfnamefont {U.}~\bibnamefont {Schneider}},\ }\href@noop {} {\bibfield
  {journal} {\bibinfo  {journal} {Phys. Rev. Lett.}\ }\textbf {\bibinfo
  {volume} {122}},\ \bibinfo {pages} {110404} (\bibinfo {year}
  {2019})}\BibitemShut {NoStop}%
\bibitem [{\citenamefont {Iyer}\ \emph {et~al.}(2013)\citenamefont {Iyer},
  \citenamefont {Oganesyan}, \citenamefont {Refael},\ and\ \citenamefont
  {Huse}}]{PhysRevB.87.134202}%
  \BibitemOpen
  \bibfield  {author} {\bibinfo {author} {\bibfnamefont {S.}~\bibnamefont
  {Iyer}}, \bibinfo {author} {\bibfnamefont {V.}~\bibnamefont {Oganesyan}},
  \bibinfo {author} {\bibfnamefont {G.}~\bibnamefont {Refael}}, \ and\ \bibinfo
  {author} {\bibfnamefont {D.~A.}\ \bibnamefont {Huse}},\ }\href {\doibase
  10.1103/PhysRevB.87.134202} {\bibfield  {journal} {\bibinfo  {journal} {Phys.
  Rev. B}\ }\textbf {\bibinfo {volume} {87}},\ \bibinfo {pages} {134202}
  (\bibinfo {year} {2013})}\BibitemShut {NoStop}%
\bibitem [{\citenamefont {L\"uschen}\ \emph {et~al.}(2017)\citenamefont
  {L\"uschen}, \citenamefont {Bordia}, \citenamefont {Scherg}, \citenamefont
  {Alet}, \citenamefont {Altman}, \citenamefont {Schneider},\ and\
  \citenamefont {Bloch}}]{PhysRevLett.119.260401}%
  \BibitemOpen
  \bibfield  {author} {\bibinfo {author} {\bibfnamefont {H.~P.}\ \bibnamefont
  {L\"uschen}}, \bibinfo {author} {\bibfnamefont {P.}~\bibnamefont {Bordia}},
  \bibinfo {author} {\bibfnamefont {S.}~\bibnamefont {Scherg}}, \bibinfo
  {author} {\bibfnamefont {F.}~\bibnamefont {Alet}}, \bibinfo {author}
  {\bibfnamefont {E.}~\bibnamefont {Altman}}, \bibinfo {author} {\bibfnamefont
  {U.}~\bibnamefont {Schneider}}, \ and\ \bibinfo {author} {\bibfnamefont
  {I.}~\bibnamefont {Bloch}},\ }\href {\doibase 10.1103/PhysRevLett.119.260401}
  {\bibfield  {journal} {\bibinfo  {journal} {Phys. Rev. Lett.}\ }\textbf
  {\bibinfo {volume} {119}},\ \bibinfo {pages} {260401} (\bibinfo {year}
  {2017})}\BibitemShut {NoStop}%
\bibitem [{\citenamefont {Bordia}\ \emph {et~al.}(2017)\citenamefont {Bordia},
  \citenamefont {L\"uschen}, \citenamefont {Scherg}, \citenamefont
  {Gopalakrishnan}, \citenamefont {Knap}, \citenamefont {Schneider},\ and\
  \citenamefont {Bloch}}]{PhysRevX.7.041047}%
  \BibitemOpen
  \bibfield  {author} {\bibinfo {author} {\bibfnamefont {P.}~\bibnamefont
  {Bordia}}, \bibinfo {author} {\bibfnamefont {H.}~\bibnamefont {L\"uschen}},
  \bibinfo {author} {\bibfnamefont {S.}~\bibnamefont {Scherg}}, \bibinfo
  {author} {\bibfnamefont {S.}~\bibnamefont {Gopalakrishnan}}, \bibinfo
  {author} {\bibfnamefont {M.}~\bibnamefont {Knap}}, \bibinfo {author}
  {\bibfnamefont {U.}~\bibnamefont {Schneider}}, \ and\ \bibinfo {author}
  {\bibfnamefont {I.}~\bibnamefont {Bloch}},\ }\href {\doibase
  10.1103/PhysRevX.7.041047} {\bibfield  {journal} {\bibinfo  {journal} {Phys.
  Rev. X}\ }\textbf {\bibinfo {volume} {7}},\ \bibinfo {pages} {041047}
  (\bibinfo {year} {2017})}\BibitemShut {NoStop}%
\bibitem [{\citenamefont {Soukoulis}\ and\ \citenamefont
  {Economou}(1982)}]{soukoulis1982localization}%
  \BibitemOpen
  \bibfield  {author} {\bibinfo {author} {\bibfnamefont {C.}~\bibnamefont
  {Soukoulis}}\ and\ \bibinfo {author} {\bibfnamefont {E.}~\bibnamefont
  {Economou}},\ }\href@noop {} {\bibfield  {journal} {\bibinfo  {journal}
  {Phys. Rev. Lett.}\ }\textbf {\bibinfo {volume} {48}},\ \bibinfo {pages}
  {1043} (\bibinfo {year} {1982})}\BibitemShut {NoStop}%
\bibitem [{\citenamefont {Kohmoto}(1983{\natexlab{b}})}]{kohmoto1983metal}%
  \BibitemOpen
  \bibfield  {author} {\bibinfo {author} {\bibfnamefont {M.}~\bibnamefont
  {Kohmoto}},\ }\href@noop {} {\bibfield  {journal} {\bibinfo  {journal} {Phys.
  Rev. Lett.}\ }\textbf {\bibinfo {volume} {51}},\ \bibinfo {pages} {1198}
  (\bibinfo {year} {1983}{\natexlab{b}})}\BibitemShut {NoStop}%
\bibitem [{\citenamefont {An}\ \emph {et~al.}(2018)\citenamefont {An},
  \citenamefont {Meier},\ and\ \citenamefont {Gadway}}]{an2018engineering}%
  \BibitemOpen
  \bibfield  {author} {\bibinfo {author} {\bibfnamefont {F.~A.}\ \bibnamefont
  {An}}, \bibinfo {author} {\bibfnamefont {E.~J.}\ \bibnamefont {Meier}}, \
  and\ \bibinfo {author} {\bibfnamefont {B.}~\bibnamefont {Gadway}},\
  }\href@noop {} {\bibfield  {journal} {\bibinfo  {journal} {Phys. Rev. X}\
  }\textbf {\bibinfo {volume} {8}},\ \bibinfo {pages} {031045} (\bibinfo {year}
  {2018})}\BibitemShut {NoStop}%
\bibitem [{\citenamefont {Bodyfelt}\ \emph {et~al.}(2014)\citenamefont
  {Bodyfelt}, \citenamefont {Leykam}, \citenamefont {Danieli}, \citenamefont
  {Yu},\ and\ \citenamefont {Flach}}]{bodyfelt2014flatbands}%
  \BibitemOpen
  \bibfield  {author} {\bibinfo {author} {\bibfnamefont {J.~D.}\ \bibnamefont
  {Bodyfelt}}, \bibinfo {author} {\bibfnamefont {D.}~\bibnamefont {Leykam}},
  \bibinfo {author} {\bibfnamefont {C.}~\bibnamefont {Danieli}}, \bibinfo
  {author} {\bibfnamefont {X.}~\bibnamefont {Yu}}, \ and\ \bibinfo {author}
  {\bibfnamefont {S.}~\bibnamefont {Flach}},\ }\href@noop {} {\bibfield
  {journal} {\bibinfo  {journal} {Phys. Rev. Lett.}\ }\textbf {\bibinfo
  {volume} {113}},\ \bibinfo {pages} {236403} (\bibinfo {year}
  {2014})}\BibitemShut {NoStop}%
\bibitem [{\citenamefont {Wang}\ \emph {et~al.}(2020)\citenamefont {Wang},
  \citenamefont {Xia}, \citenamefont {Zhang}, \citenamefont {Yao},
  \citenamefont {Chen}, \citenamefont {You}, \citenamefont {Zhou},\ and\
  \citenamefont {Liu}}]{wang2020one}%
  \BibitemOpen
  \bibfield  {author} {\bibinfo {author} {\bibfnamefont {Y.}~\bibnamefont
  {Wang}}, \bibinfo {author} {\bibfnamefont {X.}~\bibnamefont {Xia}}, \bibinfo
  {author} {\bibfnamefont {L.}~\bibnamefont {Zhang}}, \bibinfo {author}
  {\bibfnamefont {H.}~\bibnamefont {Yao}}, \bibinfo {author} {\bibfnamefont
  {S.}~\bibnamefont {Chen}}, \bibinfo {author} {\bibfnamefont {J.}~\bibnamefont
  {You}}, \bibinfo {author} {\bibfnamefont {Q.}~\bibnamefont {Zhou}}, \ and\
  \bibinfo {author} {\bibfnamefont {X.-J.}\ \bibnamefont {Liu}},\ }\href@noop
  {} {\bibfield  {journal} {\bibinfo  {journal} {Phys. Rev. Lett.}\ }\textbf
  {\bibinfo {volume} {125}},\ \bibinfo {pages} {196604} (\bibinfo {year}
  {2020})}\BibitemShut {NoStop}%
\bibitem [{\citenamefont {Yao}\ \emph {et~al.}(2019)\citenamefont {Yao},
  \citenamefont {Khoudli}, \citenamefont {Bresque},\ and\ \citenamefont
  {Sanchez-Palencia}}]{yao2019critical}%
  \BibitemOpen
  \bibfield  {author} {\bibinfo {author} {\bibfnamefont {H.}~\bibnamefont
  {Yao}}, \bibinfo {author} {\bibfnamefont {H.}~\bibnamefont {Khoudli}},
  \bibinfo {author} {\bibfnamefont {L.}~\bibnamefont {Bresque}}, \ and\
  \bibinfo {author} {\bibfnamefont {L.}~\bibnamefont {Sanchez-Palencia}},\
  }\href@noop {} {\bibfield  {journal} {\bibinfo  {journal} {Phys. Rev. Lett.}\
  }\textbf {\bibinfo {volume} {123}},\ \bibinfo {pages} {070405} (\bibinfo
  {year} {2019})}\BibitemShut {NoStop}%
\bibitem [{\citenamefont {Deng}\ \emph {et~al.}(2019)\citenamefont {Deng},
  \citenamefont {Ray}, \citenamefont {Sinha}, \citenamefont {Shlyapnikov},\
  and\ \citenamefont {Santos}}]{deng2019one}%
  \BibitemOpen
  \bibfield  {author} {\bibinfo {author} {\bibfnamefont {X.}~\bibnamefont
  {Deng}}, \bibinfo {author} {\bibfnamefont {S.}~\bibnamefont {Ray}}, \bibinfo
  {author} {\bibfnamefont {S.}~\bibnamefont {Sinha}}, \bibinfo {author}
  {\bibfnamefont {G.}~\bibnamefont {Shlyapnikov}}, \ and\ \bibinfo {author}
  {\bibfnamefont {L.}~\bibnamefont {Santos}},\ }\href@noop {} {\bibfield
  {journal} {\bibinfo  {journal} {Phys. Rev. Lett.}\ }\textbf {\bibinfo
  {volume} {123}},\ \bibinfo {pages} {025301} (\bibinfo {year}
  {2019})}\BibitemShut {NoStop}%
\bibitem [{\citenamefont {Das~Sarma}\ \emph {et~al.}(1990)\citenamefont
  {Das~Sarma}, \citenamefont {He},\ and\ \citenamefont
  {Xie}}]{PhysRevB.41.5544}%
  \BibitemOpen
  \bibfield  {author} {\bibinfo {author} {\bibfnamefont {S.}~\bibnamefont
  {Das~Sarma}}, \bibinfo {author} {\bibfnamefont {S.}~\bibnamefont {He}}, \
  and\ \bibinfo {author} {\bibfnamefont {X.~C.}\ \bibnamefont {Xie}},\ }\href
  {\doibase 10.1103/PhysRevB.41.5544} {\bibfield  {journal} {\bibinfo
  {journal} {Phys. Rev. B}\ }\textbf {\bibinfo {volume} {41}},\ \bibinfo
  {pages} {5544} (\bibinfo {year} {1990})}\BibitemShut {NoStop}%
\bibitem [{\citenamefont {Biddle}\ \emph {et~al.}(2009)\citenamefont {Biddle},
  \citenamefont {Wang}, \citenamefont {Priour},\ and\ \citenamefont
  {Das~Sarma}}]{PhysRevA.80.021603}%
  \BibitemOpen
  \bibfield  {author} {\bibinfo {author} {\bibfnamefont {J.}~\bibnamefont
  {Biddle}}, \bibinfo {author} {\bibfnamefont {B.}~\bibnamefont {Wang}},
  \bibinfo {author} {\bibfnamefont {D.~J.}\ \bibnamefont {Priour}}, \ and\
  \bibinfo {author} {\bibfnamefont {S.}~\bibnamefont {Das~Sarma}},\ }\href
  {\doibase 10.1103/PhysRevA.80.021603} {\bibfield  {journal} {\bibinfo
  {journal} {Phys. Rev. A}\ }\textbf {\bibinfo {volume} {80}},\ \bibinfo
  {pages} {021603} (\bibinfo {year} {2009})}\BibitemShut {NoStop}%
\bibitem [{\citenamefont {Biddle}\ and\ \citenamefont
  {Das~Sarma}(2010)}]{PhysRevLett.104.070601}%
  \BibitemOpen
  \bibfield  {author} {\bibinfo {author} {\bibfnamefont {J.}~\bibnamefont
  {Biddle}}\ and\ \bibinfo {author} {\bibfnamefont {S.}~\bibnamefont
  {Das~Sarma}},\ }\href {\doibase 10.1103/PhysRevLett.104.070601} {\bibfield
  {journal} {\bibinfo  {journal} {Phys. Rev. Lett.}\ }\textbf {\bibinfo
  {volume} {104}},\ \bibinfo {pages} {070601} (\bibinfo {year}
  {2010})}\BibitemShut {NoStop}%
\bibitem [{\citenamefont {Biddle}\ \emph {et~al.}(2011)\citenamefont {Biddle},
  \citenamefont {Priour}, \citenamefont {Wang},\ and\ \citenamefont
  {Das~Sarma}}]{PhysRevB.83.075105}%
  \BibitemOpen
  \bibfield  {author} {\bibinfo {author} {\bibfnamefont {J.}~\bibnamefont
  {Biddle}}, \bibinfo {author} {\bibfnamefont {D.~J.}\ \bibnamefont {Priour}},
  \bibinfo {author} {\bibfnamefont {B.}~\bibnamefont {Wang}}, \ and\ \bibinfo
  {author} {\bibfnamefont {S.}~\bibnamefont {Das~Sarma}},\ }\href {\doibase
  10.1103/PhysRevB.83.075105} {\bibfield  {journal} {\bibinfo  {journal} {Phys.
  Rev. B}\ }\textbf {\bibinfo {volume} {83}},\ \bibinfo {pages} {075105}
  (\bibinfo {year} {2011})}\BibitemShut {NoStop}%
\bibitem [{\citenamefont {Ganeshan}\ \emph {et~al.}(2015)\citenamefont
  {Ganeshan}, \citenamefont {Pixley},\ and\ \citenamefont
  {Das~Sarma}}]{PhysRevLett.114.146601}%
  \BibitemOpen
  \bibfield  {author} {\bibinfo {author} {\bibfnamefont {S.}~\bibnamefont
  {Ganeshan}}, \bibinfo {author} {\bibfnamefont {J.~H.}\ \bibnamefont
  {Pixley}}, \ and\ \bibinfo {author} {\bibfnamefont {S.}~\bibnamefont
  {Das~Sarma}},\ }\href {\doibase 10.1103/PhysRevLett.114.146601} {\bibfield
  {journal} {\bibinfo  {journal} {Phys. Rev. Lett.}\ }\textbf {\bibinfo
  {volume} {114}},\ \bibinfo {pages} {146601} (\bibinfo {year}
  {2015})}\BibitemShut {NoStop}%
\bibitem [{\citenamefont {An}\ \emph {et~al.}(2021)\citenamefont {An},
  \citenamefont {Padavi\ifmmode~\acute{c}\else \'{c}\fi{}}, \citenamefont
  {Meier}, \citenamefont {Hegde}, \citenamefont {Ganeshan}, \citenamefont
  {Pixley}, \citenamefont {Vishveshwara},\ and\ \citenamefont
  {Gadway}}]{PhysRevLett.126.040603}%
  \BibitemOpen
  \bibfield  {author} {\bibinfo {author} {\bibfnamefont {F.~A.}\ \bibnamefont
  {An}}, \bibinfo {author} {\bibfnamefont {K.}~\bibnamefont
  {Padavi\ifmmode~\acute{c}\else \'{c}\fi{}}}, \bibinfo {author} {\bibfnamefont
  {E.~J.}\ \bibnamefont {Meier}}, \bibinfo {author} {\bibfnamefont
  {S.}~\bibnamefont {Hegde}}, \bibinfo {author} {\bibfnamefont
  {S.}~\bibnamefont {Ganeshan}}, \bibinfo {author} {\bibfnamefont {J.~H.}\
  \bibnamefont {Pixley}}, \bibinfo {author} {\bibfnamefont {S.}~\bibnamefont
  {Vishveshwara}}, \ and\ \bibinfo {author} {\bibfnamefont {B.}~\bibnamefont
  {Gadway}},\ }\href {\doibase 10.1103/PhysRevLett.126.040603} {\bibfield
  {journal} {\bibinfo  {journal} {Phys. Rev. Lett.}\ }\textbf {\bibinfo
  {volume} {126}},\ \bibinfo {pages} {040603} (\bibinfo {year}
  {2021})}\BibitemShut {NoStop}%
\bibitem [{\citenamefont {Roy}\ \emph {et~al.}(2021{\natexlab{b}})\citenamefont
  {Roy}, \citenamefont {Mukerjee},\ and\ \citenamefont
  {Kulkarni}}]{PhysRevB.103.184203}%
  \BibitemOpen
  \bibfield  {author} {\bibinfo {author} {\bibfnamefont {S.}~\bibnamefont
  {Roy}}, \bibinfo {author} {\bibfnamefont {S.}~\bibnamefont {Mukerjee}}, \
  and\ \bibinfo {author} {\bibfnamefont {M.}~\bibnamefont {Kulkarni}},\ }\href
  {\doibase 10.1103/PhysRevB.103.184203} {\bibfield  {journal} {\bibinfo
  {journal} {Phys. Rev. B}\ }\textbf {\bibinfo {volume} {103}},\ \bibinfo
  {pages} {184203} (\bibinfo {year} {2021}{\natexlab{b}})}\BibitemShut
  {NoStop}%
\bibitem [{\citenamefont {Siebesma}\ and\ \citenamefont
  {Pietronero}(1987)}]{Siebesma_1987}%
  \BibitemOpen
  \bibfield  {author} {\bibinfo {author} {\bibfnamefont {A.~P.}\ \bibnamefont
  {Siebesma}}\ and\ \bibinfo {author} {\bibfnamefont {L.}~\bibnamefont
  {Pietronero}},\ }\href {\doibase 10.1209/0295-5075/4/5/014} {\bibfield
  {journal} {\bibinfo  {journal} {Europhysics Letters ({EPL})}\ }\textbf
  {\bibinfo {volume} {4}},\ \bibinfo {pages} {597} (\bibinfo {year}
  {1987})}\BibitemShut {NoStop}%
\bibitem [{\citenamefont {Kohmoto}\ \emph {et~al.}(1983)\citenamefont
  {Kohmoto}, \citenamefont {Kadanoff},\ and\ \citenamefont
  {Tang}}]{PhysRevLett.50.1870}%
  \BibitemOpen
  \bibfield  {author} {\bibinfo {author} {\bibfnamefont {M.}~\bibnamefont
  {Kohmoto}}, \bibinfo {author} {\bibfnamefont {L.~P.}\ \bibnamefont
  {Kadanoff}}, \ and\ \bibinfo {author} {\bibfnamefont {C.}~\bibnamefont
  {Tang}},\ }\href {\doibase 10.1103/PhysRevLett.50.1870} {\bibfield  {journal}
  {\bibinfo  {journal} {Phys. Rev. Lett.}\ }\textbf {\bibinfo {volume} {50}},\
  \bibinfo {pages} {1870} (\bibinfo {year} {1983})}\BibitemShut {NoStop}%
\bibitem [{\citenamefont {Szab{\'o}}\ and\ \citenamefont
  {Schneider}(2018)}]{szabo2018non}%
  \BibitemOpen
  \bibfield  {author} {\bibinfo {author} {\bibfnamefont {A.}~\bibnamefont
  {Szab{\'o}}}\ and\ \bibinfo {author} {\bibfnamefont {U.}~\bibnamefont
  {Schneider}},\ }\href@noop {} {\bibfield  {journal} {\bibinfo  {journal}
  {Phys. Rev. B}\ }\textbf {\bibinfo {volume} {98}},\ \bibinfo {pages} {134201}
  (\bibinfo {year} {2018})}\BibitemShut {NoStop}%
\bibitem [{\citenamefont {Han}\ \emph {et~al.}(1994)\citenamefont {Han},
  \citenamefont {Thouless}, \citenamefont {Hiramoto},\ and\ \citenamefont
  {Kohmoto}}]{PhysRevB.50.11365}%
  \BibitemOpen
  \bibfield  {author} {\bibinfo {author} {\bibfnamefont {J.~H.}\ \bibnamefont
  {Han}}, \bibinfo {author} {\bibfnamefont {D.~J.}\ \bibnamefont {Thouless}},
  \bibinfo {author} {\bibfnamefont {H.}~\bibnamefont {Hiramoto}}, \ and\
  \bibinfo {author} {\bibfnamefont {M.}~\bibnamefont {Kohmoto}},\ }\href
  {\doibase 10.1103/PhysRevB.50.11365} {\bibfield  {journal} {\bibinfo
  {journal} {Phys. Rev. B}\ }\textbf {\bibinfo {volume} {50}},\ \bibinfo
  {pages} {11365} (\bibinfo {year} {1994})}\BibitemShut {NoStop}%
\bibitem [{\citenamefont {Li}\ \emph {et~al.}(2017)\citenamefont {Li},
  \citenamefont {Li},\ and\ \citenamefont {Das~Sarma}}]{PhysRevB.96.085119}%
  \BibitemOpen
  \bibfield  {author} {\bibinfo {author} {\bibfnamefont {X.}~\bibnamefont
  {Li}}, \bibinfo {author} {\bibfnamefont {X.}~\bibnamefont {Li}}, \ and\
  \bibinfo {author} {\bibfnamefont {S.}~\bibnamefont {Das~Sarma}},\ }\href
  {\doibase 10.1103/PhysRevB.96.085119} {\bibfield  {journal} {\bibinfo
  {journal} {Phys. Rev. B}\ }\textbf {\bibinfo {volume} {96}},\ \bibinfo
  {pages} {085119} (\bibinfo {year} {2017})}\BibitemShut {NoStop}%
\bibitem [{\citenamefont {Roy}\ \emph {et~al.}(2018)\citenamefont {Roy},
  \citenamefont {Khaymovich}, \citenamefont {Das},\ and\ \citenamefont
  {Moessner}}]{10.21468/SciPostPhys.4.5.025}%
  \BibitemOpen
  \bibfield  {author} {\bibinfo {author} {\bibfnamefont {S.}~\bibnamefont
  {Roy}}, \bibinfo {author} {\bibfnamefont {I.~M.}\ \bibnamefont {Khaymovich}},
  \bibinfo {author} {\bibfnamefont {A.}~\bibnamefont {Das}}, \ and\ \bibinfo
  {author} {\bibfnamefont {R.}~\bibnamefont {Moessner}},\ }\href {\doibase
  10.21468/SciPostPhys.4.5.025} {\bibfield  {journal} {\bibinfo  {journal}
  {SciPost Phys.}\ }\textbf {\bibinfo {volume} {4}},\ \bibinfo {pages} {25}
  (\bibinfo {year} {2018})}\BibitemShut {NoStop}%
\bibitem [{\citenamefont {Tang}\ and\ \citenamefont
  {Kohmoto}(1986)}]{PhysRevB.34.2041}%
  \BibitemOpen
  \bibfield  {author} {\bibinfo {author} {\bibfnamefont {C.}~\bibnamefont
  {Tang}}\ and\ \bibinfo {author} {\bibfnamefont {M.}~\bibnamefont {Kohmoto}},\
  }\href {\doibase 10.1103/PhysRevB.34.2041} {\bibfield  {journal} {\bibinfo
  {journal} {Phys. Rev. B}\ }\textbf {\bibinfo {volume} {34}},\ \bibinfo
  {pages} {2041} (\bibinfo {year} {1986})}\BibitemShut {NoStop}%
\bibitem [{\citenamefont {Kohmoto}\ \emph {et~al.}(1987)\citenamefont
  {Kohmoto}, \citenamefont {Sutherland},\ and\ \citenamefont
  {Tang}}]{PhysRevB.35.1020}%
  \BibitemOpen
  \bibfield  {author} {\bibinfo {author} {\bibfnamefont {M.}~\bibnamefont
  {Kohmoto}}, \bibinfo {author} {\bibfnamefont {B.}~\bibnamefont {Sutherland}},
  \ and\ \bibinfo {author} {\bibfnamefont {C.}~\bibnamefont {Tang}},\ }\href
  {\doibase 10.1103/PhysRevB.35.1020} {\bibfield  {journal} {\bibinfo
  {journal} {Phys. Rev. B}\ }\textbf {\bibinfo {volume} {35}},\ \bibinfo
  {pages} {1020} (\bibinfo {year} {1987})}\BibitemShut {NoStop}%
\bibitem [{\citenamefont {Hashimoto}\ \emph {et~al.}(1992)\citenamefont
  {Hashimoto}, \citenamefont {Niizeki},\ and\ \citenamefont
  {Okabe}}]{Hashimoto_1992}%
  \BibitemOpen
  \bibfield  {author} {\bibinfo {author} {\bibfnamefont {Y.}~\bibnamefont
  {Hashimoto}}, \bibinfo {author} {\bibfnamefont {K.}~\bibnamefont {Niizeki}},
  \ and\ \bibinfo {author} {\bibfnamefont {Y.}~\bibnamefont {Okabe}},\ }\href
  {\doibase 10.1088/0305-4470/25/20/005} {\bibfield  {journal} {\bibinfo
  {journal} {Journal of Physics A: Mathematical and General}\ }\textbf
  {\bibinfo {volume} {25}},\ \bibinfo {pages} {5211} (\bibinfo {year}
  {1992})}\BibitemShut {NoStop}%
\bibitem [{\citenamefont {Evers}\ and\ \citenamefont
  {Mirlin}(2008)}]{RevModPhys.80.1355}%
  \BibitemOpen
  \bibfield  {author} {\bibinfo {author} {\bibfnamefont {F.}~\bibnamefont
  {Evers}}\ and\ \bibinfo {author} {\bibfnamefont {A.~D.}\ \bibnamefont
  {Mirlin}},\ }\href {\doibase 10.1103/RevModPhys.80.1355} {\bibfield
  {journal} {\bibinfo  {journal} {Rev. Mod. Phys.}\ }\textbf {\bibinfo {volume}
  {80}},\ \bibinfo {pages} {1355} (\bibinfo {year} {2008})}\BibitemShut
  {NoStop}%
\bibitem [{\citenamefont {Ikezawa}\ and\ \citenamefont
  {Kohmoto}(1994)}]{ikezawa1994energy}%
  \BibitemOpen
  \bibfield  {author} {\bibinfo {author} {\bibfnamefont {K.}~\bibnamefont
  {Ikezawa}}\ and\ \bibinfo {author} {\bibfnamefont {M.}~\bibnamefont
  {Kohmoto}},\ }\href@noop {} {\bibfield  {journal} {\bibinfo  {journal}
  {Journal of the Physical Society of Japan}\ }\textbf {\bibinfo {volume}
  {63}},\ \bibinfo {pages} {2261} (\bibinfo {year} {1994})}\BibitemShut
  {NoStop}%
\end{thebibliography}%

\end{document}